\documentclass[reprint,superscriptaddress,aps,]{revtex4-1}

\usepackage{amsmath}
\usepackage{amssymb}
\usepackage{amsfonts}
\usepackage{gensymb}
\usepackage{color}
\usepackage{graphicx}
\usepackage{dcolumn}
\usepackage{bm}
\usepackage{times}
\usepackage[colorlinks=true,linkcolor=blue,citecolor=blue]{hyperref}

\begin{document}

\title{Crystal growth, microstructure and physical properties of SrMnSb$_2$}

\author{Yong Liu}
\email{yongliu31@outlook.com}

\author{Tao Ma}

\author{Lin Zhou}

\author{Warren E. Straszheim}

\author{Farhan Islam}

\author{Brandt A. Jensen}

\affiliation{Division of Materials Sciences and Engineering, Ames Laboratory, U.S. DOE, Ames, Iowa 50011, USA}

\author{Wei Tian}

\affiliation{Neutron Scattering Division, Oak Ridge National Laboratory, Oak Ridge, Tennessee 37831, USA}

\author{Thomas Heitmann}

\affiliation{The Missouri Research Reactor, University of Missouri, Columbia, Missouri 65211, USA}

\author{R. A. Rosenberg}

\affiliation{Advanced Photon Source, Argonne National Laboratory, Argonne, Illinois 60439, USA}

\author{J. M. Wilde}

\author{Bing Li}

\author{Andreas Kreyssig}

\author{Alan I. Goldman}

\author{B. G. Ueland}

\author{Robert J. McQueeney}

\author{David Vaknin}
\email{vaknin@ameslab.gov}

\affiliation{Division of Materials Sciences and Engineering, Ames Laboratory, U.S. DOE, Ames, Iowa 50011, USA}
\affiliation{Department of Physics and Astronomy, Iowa State University, Ames, Iowa 50011, USA}

\date{\today}

\begin{abstract}

We report on the crystal and magnetic structures, magnetic, and transport properties of SrMnSb$_2$ single crystals grown by the self-flux method. Magnetic susceptibility measurements reveal an antiferromagnetic (AFM) transition at $T_{\rm N} = 295(3)$ K. Above $T_{\rm N}$, the susceptibility slightly increases and forms a broad peak at $T \sim 420$ K, which is a typical feature of two-dimensional  magnetic systems. Neutron diffraction measurements on single crystals confirm the previously reported C-type AFM structure below $T_{\rm N}$. Both de Haas-van Alphen (dHvA) and Shubnikov-de Haas (SdH) effects are observed in SrMnSb$_2$ single crystals. Analysis of the oscillatory component by a Fourier transform shows that the prominent frequencies obtained by the two different techniques are practically the same within error regardless of sample size or saturated magnetic moment. Transmission electron microscopy (TEM) reveals the existence of stacking faults in the crystals, which result from a horizontal shift of Sb atomic layers suggesting possible ordering of Sb vacancies in the crystals. Increase of temperature in susceptibility  measurements leads to the formation of a strong peak at $T \sim {570}$ K that upon cooling under magnetic field the susceptibility shows a ferromagnetic transition at $T_{\rm C} \sim 580$ K. Neutron powder diffraction on crushed single-crystals does not support an FM phase above $T_{\rm N}$. Furthermore, X-ray magnetic circular dichroism (XMCD) measurements of a single crystal at the $L_{2,3}$ edge of Mn shows a signal due to induced canting of AFM moments by the applied magnetic field. All evidence strongly suggests that a chemical transformation at the surface of single crystals occurs above 500 K concurrently producing a minute amount of ferromagnetic  impurity phase.

\end{abstract}

\maketitle

\section{Introduction}

The discovery of two dimensional (2D) graphene \cite{Geim2007} has advanced the interest in the search for other 2D or three dimensional (3D) materials analogs with linearly dispersing bands that cross at the Fermi energy \cite{Wehling2014}. In these materials (now referred to as Dirac materials) \cite{Wehling2014}, low-energy fermionic excitations behave as massless Dirac quasiparticles. In the case of broken time-reversal or space-inversion symmetry, Dirac materials may further evolve into Weyl semimetals (WSMs) \cite{Nielsen1981,Wan2011}. In WSMs, the crossing points of the nondegenerate valence and conduction bands are referred to as Weyl nodes which appear in pairs in the Brillouin zone with opposite chirality, left-handed or right-handed \cite{Nielsen1981}. The Weyl node pairs are separated in momentum space and appear to be connected by a Fermi arc when projected onto the surface of the material \cite{Nielsen1981,Wan2011}. Experimentally, the Weyl nodes and Fermi arc have been observed in the broken-inversion-symmetry TaAs \cite{Huang2015,Xu2015,Lv2015,Yang2015} and NbAs \cite{Xu2015b} by angle-resolved photoemission spectroscopy measurements (ARPES).

While there are several examples of WSMs with broken inversion symmetry, there are fewer confirmed materials with broken time-reversal symmetry. In this respect, the observation of highly anisotropic Dirac fermions in a Bi square net in SrMnBi$_2$ \cite{Park2011}, $A$MnBi$_2$ and $A$MnSb$_2$ ($A$ = Ca, Sr, Ba, Yb, Eu) materials with magnetic moments on the Mn or Eu sites (Yb$^{2+}$ are non-magnetic ions) have attracted much attention \cite{Wang2012,May2014,Masuda2016,Liu2017b,Yi2017}. ARPES measurements have provided direct evidence for the existence of strongly anisotropic Dirac cones in SrMnBi$_2$ and CaMnBi$_2$ \cite{Feng2014}. And first-principles calculations of $A$MnSb$_2$ ($A$ = Sr, Ba; replacing Bi with Sb) have indicated Dirac fermionic behavior could be realized in BaMnSb$_2$ \cite{Farhan2014}.

Subsequent transport and magnetization measurements and relativistic first-principles calculations have suggested that BaMnSb$_2$ is a 3D WSM by virtue of weak ferromagnetism due to canted Mn moments  in the antiferromagnetic (AFM) structure \cite{Huang2017}. The reported canting leads to a ferromagnetic (FM) component which breaks the time-reversal symmetry in the crystal and establishes WSM behavior. However, neutron scattering measurements found that the magnetic ground-state of stoichiometric BaMnSb$_2$ has G-type AFM order with no clear evidence to support the existence of any FM canting \cite{Liu2016}. In fact, symmetry analysis in Ref. \cite{Liu2016} shows that FM canting of the G-type order is not allowed in host tetragonal structures.

The orthorhombic distortion in SrMnSb$_2$ can host a FM canting of the AFM structure based on symmetry considerations. Its crystal structure consists of MnSb layers stacked along \textit{\textbf{a}} axis, as shown in Fig. \ref{Fig:Crystals}(c). Interestingly, FM behavior was reported in Mn- and Sr-deficient Sr$_{1-y}$Mn$_{1-z}$Sb$_2$ ($y, z<0.1$) single crystals \cite{Liu2017}. Previous neutron scattering measurements also indicated that this material exhibits an unusual high-temperature FM ordered phase for $304<T<565$ K which evolves into canted AFM order with a small in-plane FM component for $T<304$ K \cite{Liu2017}. In addition, it was reported that variations from stoichiometry in Mn and/or Sr content have a strong effect on the magnetic properties of Sr$_{1-y}$Mn$_{1-z}$Sb$_2$. In particular, it is claimed that samples with $y \sim 0.08$ and $z \sim 0.02$ display a larger saturated moment, while weaker FM behavior occurs in samples with enhanced Mn deficiency $y \sim 0.01-0.04$, $z \sim 0.04-0.1$.

It should be noted that, recent ARPES, transport measurements, and density functional theory (DFT) and tight binding calculations show that the electronic bands are gapped above the Fermi energy for SrMnSb$_2$, \emph{suggesting trivial topology} \cite{Ramankutty2018}. The extracted Berry phase is zero, \emph{indicating the non-topological character} of the transport in SrMnSb$_2$ \cite{Ramankutty2018}, whereas the non-trivial $\pi$ Berry phase was observed in Type A crystals in Ref. \cite{Liu2017}. This makes the thorough investigation of SrMnSb$_2$ of paramount importance for the advancement of the science of topological semimetals.

Here, we report on the growth of SrMnSb$_2$ single crystals and examine their magnetic, transport, and structural stability using various experimental techniques. In particular, we focus on the properties of single-crystals with little or no net ferromagnetic response in order to study the intrinsic properties of the purely AFM state. Neutron diffraction and magnetic susceptibility measurements indicate quasi-2D AFM with strong 2D correlations persisting up to 500 K. TEM measurements indicate the presence of stacking faults, as might be expected for a layered system. However, no other significant defect structures or inclusions are observed. X-ray magnetic circular dichroism (XMCD) measurements find no intrinsic FM signal. Quantum oscillations are consistent with high mobility 2D electron transport reported for SrMnSb$_2$. By comparison to samples with a significant FM component, we find that all of these experimental signatures are relatively independent of the presence of ferromagnetism. Interestingly, we demonstrate that an irreversible FM component develops above 560 K, presumably due to a chemical transformation. Thus, it is likely that the origin of FM in our samples is extrinsic in origin, although we could not identify its source.

\section{Experimental Details}

Plate-like SrMnSb$_2$ single crystals were grown by a self-flux method from different starting compositions and soaking temperatures. One set of crystals was grown from a starting composition of Sr:Mn:Sb=1:1:4, referred to as Batch A. Strontium dendritic pieces, manganese and antimony powders were weighed and loaded in an alumina crucible in a glovebox under argon atmosphere. The alumina crucible was sealed in a quartz tube and installed in a vertical tube furnace. The assembly was heated up to a soaking temperature of 1073 K. The quartz ampoule was cooled down to 773 K at a 3 K/h cooling rate. Single crystals in Batch B were grown from a starting composition of Sr:Mn:Sb=1:1:2 and loaded in a boron nitride (BN) crucible. The assembly was heated up to 1323 K for a dwell time of 6 hours and slowly cooled down to 773 K at the same rate of 3 K/h. The furnace was then shut down and the temperature dropped below 373 K in 4 hours. 

\begin{figure}
\centering
\includegraphics[width=0.8\linewidth]{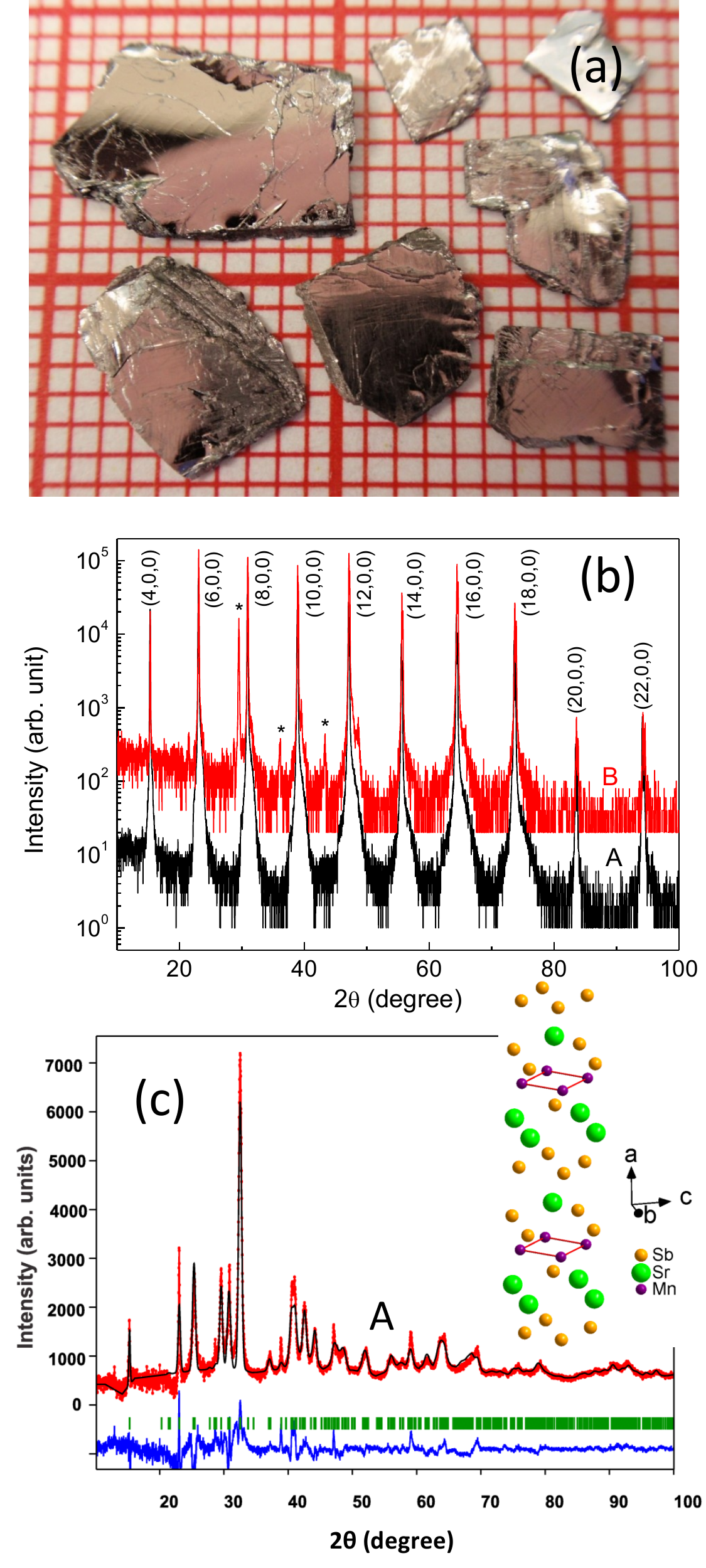}
\caption{(Color online) (a) Photograph of SrMnSb$_2$ single crystals cleaved from the growth of Batch A on a square graph paper with 1 mm grid. (b) XRD patterns of single crystals from Batches A and B showing the intensity versus scattering angle $2\theta$ (using Cu-K$\alpha$ X-ray source) of the ($h$,0,0) reflections up to the 22th order for both batches of crystals. The intensity on log scale shows weak extra peaks at a level of $\leq 1\%$ of the main Bragg peaks of the main phase for the crystals from Batch B, as indicated by the asterisk. (c) XRD of crushed single crystal (Batch A) including a refinement of the $Pnma$ structure using the \textsc{FullProf} suite. Inset shows the crystal structure of SrMnSb$_2$.} 
\label{Fig:Crystals}
\end{figure}

Images of large plate-like single crystals cleaved from Batch A are shown in Fig. \ref{Fig:Crystals}(a). It should be pointed out that Sb has a melting point of 904 K, and there is an eutectic point at 843 K for the composition of 81 at.\% Sb and 19 at.\% Mn in Mn-Sb binary phase diagram \cite{Okamoto1990}. By choosing the starting Sr:Mn:Sb=1:1:4 molar ratio, one may expect a high yield of large single crystals grown over a relatively broad temperature range. Indeed, the starting Sr:Mn:Sb=1:1:4 molar ratio yields much larger single crystals than those grown from the starting Sr:Mn:Sb=1:1:2 molar ratio.

To determine crystal quality, X-ray diffraction (XRD) measurements were performed on the single crystals using a Bruker D8 Advance Powder Diffractometer using Cu K$\alpha$ radiation. Figure \ref{Fig:Crystals}(b) shows ($h$,0,0) XRD patterns (shown on log scale intensity) of single crystals from Batches A and B, respectively. Whereas the ($h$,0,0) pattern for Batch-A crystals is void of any extra peaks, Batch-B crystals show few extra peaks at a level of less than 1\% in intensity compared to the main ($h$,0,0) peaks. The impurity phase with these extra peaks is consistent with the crystal structure of Sr$_2$Mn$_3$Sb$_2$O$_2$ \cite{Brock1996a,Brock1996b}. As we discuss below, SrMnSb$_2$ crystals have a strong tendency to react with air and deteriorate at the surface, which could be the source of the oxide formation. From the XRD patterns shown in Fig. \ref{Fig:Crystals}(b), we extract the lattice parameter $a=23.051(5)$ {\AA} for both types of growths.

To further characterize the phase purity we crushed a few crystals from Batches A and B in an inert glovebox and loaded the powder in an air-tight sample holder covered by Kapton tape and conducted X-ray powder diffraction on the same Bruker D8 diffractometer. Figure \ref{Fig:Crystals}(c) shows XRD of Batch A sample (after subtracting the empty Kapton holder signal) along with Rietveld refinements using \textsc{FullProf} software \cite{FullProf} of the $Pnma$ space group. The refinement yields the following lattice constants $a = 23.05(5)$,  $ b = 4.3816(6)$, $c = 4.4254(6)$ {\AA}. These results are in good agreement with previous reports\cite{Liu2017,Brechtel1981}.

Magnetization measurements were performed using a Physical Property Measurement System (PPMS, Quantum Design) equipped with a vibrating sample magnetometer (VSM). For temperature dependent magnetization measurements, the samples were cooled down to the desired temperature either with or without application of a magnetic field, termed as ZFC and FC, respectively. The temperature dependent magnetization data were then collected upon warming at 2 K/min under a fixed field. The magnetic field \textbf{H} was applied either parallel (\textbf{H}$\perp$\textit{\textbf{a}} axis) or perpendicular to the plate (\textbf{H}$||$\textit{\textbf{a}} axis). For the high-temperature ($T >$ 300 K) magnetization measurements, crystals were glued to a heater stick using Zircar Cement. After the cement dried, copper foil was wrapped around the sample to provide a good isothermal region. We note that for the measurements from 2 to 340 K, the sample is placed in a quartz holder and kept at $p \approx 10$ Torr, and for the temperature range $300<T<650$ K, the sample was glued to a heater stick and kept under vacuum ($10^{-5}$ Torr). The magnetization data were collected upon warming under a field to 650 K. After reaching 650 K the sample was cooled down to 300 K under field.

The longitudinal resistivity $\rho_{xx}$ was measured in a PPMS using a conventional four-probe method with an applied current in the \textit{\textbf{bc}} plane (i.e., in the basal layer). Hall (transverse) resistivity $\rho_{xy}$ was measured as a function of magnetic field at a fixed temperature. Four probe contacts were made by soldering thin gold wires to the single crystals. The Hall coefficient is extracted from the slope of the linear field dependence of Hall resistivity by sweeping the magnetic field. A driven current of 5 mA at 19 Hz was used in the Hall effect measurements.

Single-crystal neutron diffraction experiments were carried out on the HB-1A triple axis spectrometer located at the High Flux Isotope Reactor at Oak Ridge National Laboratory. The HB-1A spectrometer operates with a fixed incident energy of $E_i=14.64$ meV using a double pyrolytic graphite (PG) monochromator system. PG filters were placed before and after the second monochromator to reduce higher order contamination in the incident beam achieving a ratio $\frac{I_\lambda}{2}:I_\lambda: \approx 10^{-4}$. The SrMnSb$_2$ crystal was mounted with the (0,$k$,$l$) and ($h$,$k$,0) planes in the scattering plane.

Powder neutron diffraction experiments on crushed SrMnSb$_2$ crystals were conducted on the PSD diffractometer at the University of Missouri Research Reactor ($\lambda = 1.485$ {\AA}) at the following temperature sequence: 300, 400, 500, and 600 K, and back to 300 K. Approximately 1.5 g of powder was loaded in a helium atmosphere and mounted in a closed-cycle furnace-refrigerator on the PSD diffractometer. Rietveld refinements of the neutron and X-ray diffraction data were carried out using the \textsc{FullProf} software suite \cite{FullProf}.

X-ray absorption (XAS) and X-ray magnetic circular dichroism (XMCD) spectra at the Mn $L_{3}$ and $L_{2}$  absorption edges were recorded for a thin ($\approx 100~\mu$m) single-crystal sample of SrMnSb$_{2}$ using end station $4$-ID-C beam line at the Advanced Photon Source, Argonne National Laboratory. The samples were stored under inert gas in order to minimize oxidation prior to being mounted on carbon tape and inserted into the bore of a horizontal superconducting magnet with an incorporated helium-flow cryostat. The single-crystal sample was cleaved immediately prior to mounting, and was first aligned with its \textit{\textbf{a}} axis along the horizontal X-ray beam.  It was then rotated by $45 \degree$ around its vertical axis in order to achieve a nonzero value of magnetic field \textbf{H} within the \textit{\textbf{bc}} plane.  XAS and XMCD signals were recorded for the single-crystal and powder samples as a function of \textbf{H} up to $H=5.5$ T at a temperature of $T=230$~K. Total electron yield (TEY) and total fluorescence yield (TFY) data were recorded \cite{Lang2003}. Data taken for the powder sample of SrMnSb$_2$ yielded results similar to those for the single-crystal sample.

The composition of the crystals was determined by using an Oxford Instruments energy-dispersive X-ray spectroscopy (EDS) system on a Thermo Scientific Teneo scanning electron microscopy (SEM). Transmission electron microscopy (TEM) samples were prepared using a Thermo Scientific Helios NanoLab G3 UC with EasyLift micromanipulator and MultiChem Gas Injection System. A Thermo Scientific Titan Themis 300 probe aberration-corrected scanning transmission electron microscope (AC-STEM), operated under an accelerating voltage of 200 kV and equipped with a Bruker$^{\rm TM}$ Super-X energy-dispersive X-ray spectroscopy (EDS) detector, was employed for TEM observation. The high-angle annular dark-field (HAADF) images were collected in a range of 99-200 mrad; and the annular bright-field (ABF) images were collected in a range of 10-22 mrad. The element mapping was performed to reveal the distribution of all involved elements in the field of view by using EDS built in AC-STEM. To preserve the surface oxidation layer, a 200 nm thick carbon-protective layer was deposited using e-beam, followed by the deposition of a 3 $\mu$m thick carbon-protective layer using ion beam. EDS measurements on both types of crystals show a uniform composition: Sr at $0.91\pm 0.02$ Mn at $0.95\pm 0.02$ and Sb at 2. These results are close to those reported in Ref. \cite{Liu2017}.

\section{Results and Discussion}
\subsection{Magnetic Susceptibility and de Haas-van Alphen effect}

\begin{figure}
\centering
\includegraphics[width=1.0\linewidth]{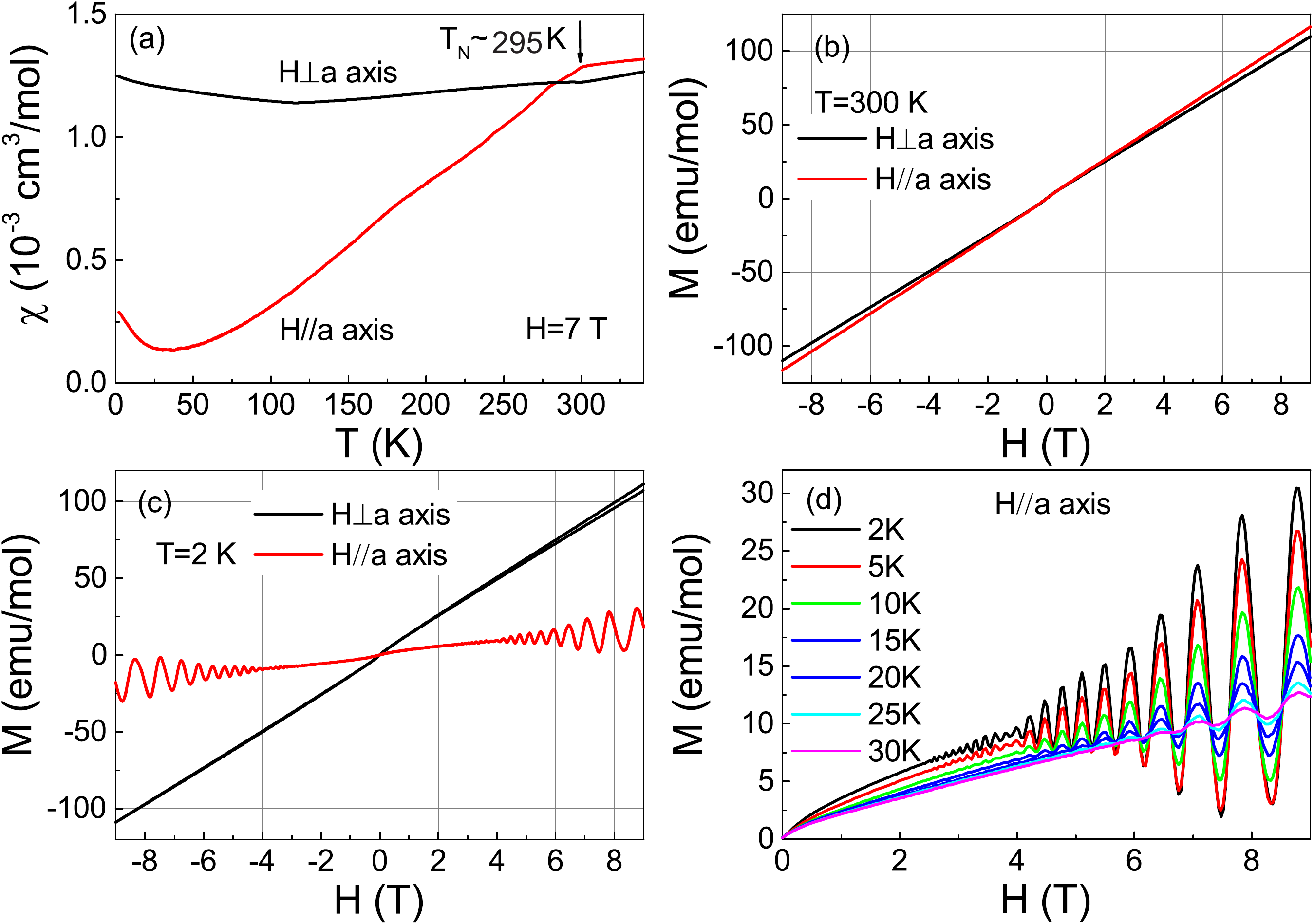}
\caption{(Color online)  (a) Temperature dependence of magnetic susceptibility $\chi(T)$ in both \textbf{H}$\perp$\textit{\textbf{a}} axis and \textbf{H}$||$\textit{\textbf{a}} axis configurations for SrMnSb$_2$ crystals from Batch A. Arrow indicates the antiferromagnetic (AFM) transition at $T_{N}=295$ K. (b) Magnetization hysteresis loops (MHLs) at $T=300$ K. Linear field dependence is observed. (c) MHLs at $T=2$ K. (d) Isothermal magnetization in \textbf{H}$||$\textit{\textbf{a}} axis at $T=2, 5, 10, 15, 20, 25, 30$ K.} 
\label{Fig:Mag-I}
\end{figure}

\begin{figure}
\centering
\includegraphics[width=1.0\linewidth]{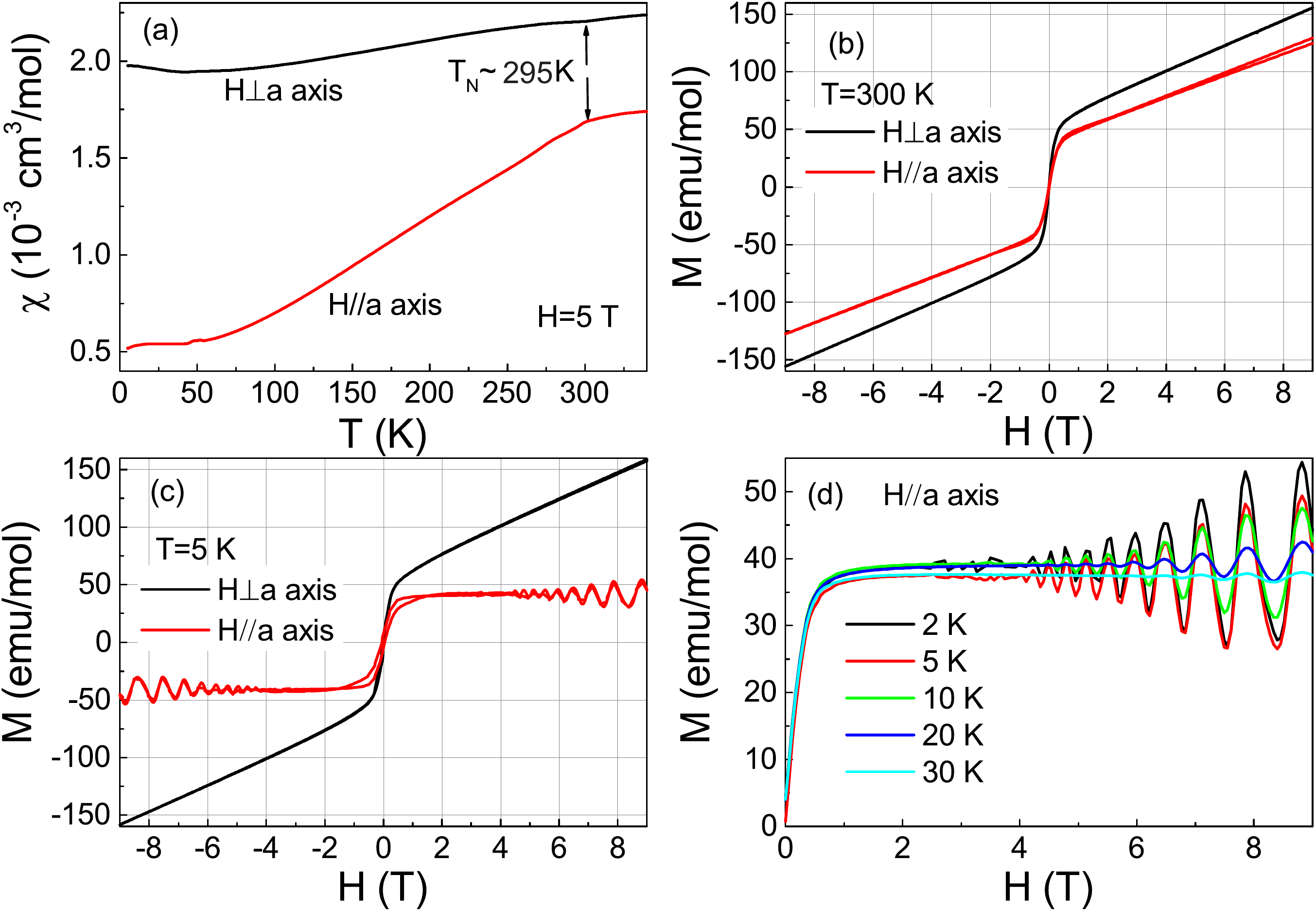}
\caption{(Color online)  (a) Temperature dependence of magnetic susceptibility $\chi(T)$ in both \textbf{H}$\perp$\textit{\textbf{a}} axis and \textbf{H}$||$\textit{\textbf{a}} axis configurations for SrMnSb$_2$ crystals from Batch B. (b) Magnetization hysteresis loops (MHLs) at $T=300$ K. (c) MHLs at $T=5$ K. (d) Isothermal magnetization in \textbf{H}$||$\textit{\textbf{a}} axis at $T=2, 5, 10, 20, 30$ K. A ferromagnetic (FM) component is observed. } 
\label{Fig:Mag-II}
\end{figure}

Figures \ref{Fig:Mag-I} and \ref{Fig:Mag-II} show susceptibility and magnetization data for SrMnSb$_2$ crystals from Batches A and B, respectively. Panel (a) of both figures shows that the magnetic susceptibilities $\chi(T) = \frac{M}{H}$ for \textbf{H}$\perp$\textit{\textbf{a}} and \textbf{H}$||$\textit{\textbf{a}} decrease with decreasing temperature below a kink that indicates the AFM transition at $T_{\rm N}$ $\approx 295$ K in both samples, with anisotropy below $T_{\rm N}$ indicating that ordered moments lie along \textit{\textbf{a}}. The two samples display similar temperature dependence and anisotropy, but differ in the magnitude of $\chi(T)$ in the two field orientations, generally indicative of a small FM contribution that in our case is more apparent in the Batch B sample.

Two features in the susceptibility indicate that both samples maintain strong 2D correlations above the 3D AFM ordering at $T_{\rm N}$. First, there is only a subtle susceptibility signature at $T_{\rm N}$ in the form of a small change in slope. Second, the susceptibility above $T_{\rm N}$ does not display Curie-Weiss behaviour, namely, $\chi(T) \propto 1/T$ above the transition, but rather increases linearly with temperature above $T_{\rm N}$ [$\chi(T) \propto T$]. This feature is typically associated with 2D behavior, where there is a broad peak in the susceptibility above $T_{\rm N}$. In fact, the extended susceptibility measurements above $T_{\rm N}$ to about 600 K confirm the increase and development of a broad peak. 

Panels (b) and (c) of Figs. \ref{Fig:Mag-I} and \ref{Fig:Mag-II} show the magnetization hysteresis loops (MHLs) for \textbf{H}$\perp$\textit{\textbf{a}} and \textbf{H}$||$\textit{\textbf{a}} slightly above $T_{\rm N}$ and  below $T_{\rm N}$ (2 and 5 K, respectively) for the two samples. Panel (d) of Figs. \ref{Fig:Mag-I} and \ref{Fig:Mag-II} shows clear de Haas-van Alphen (dHvA) oscillations that persist to about 30 K for both samples for field \textbf{H}$||$\textit{\textbf{a}} up to 9 T. The nearly linear behavior of the magnetization with \textbf{H}$\perp$\textit{\textbf{a}} of Batch A crystals is characteristic of a AFM system with moments aligned normal to the applied magnetic field, consistent with the Type C crystal data in Ref. \cite{Liu2017}. By contrast, the $M$ versus $H$ curve for the Batch B crystals seems to be a superposition of a linear curve (just like the one of Batch A crystals) and a smooth step-like function characteristic of a saturated hysteresis  curve of a ferromagnet. It should be noted that the AFM transition in Type A crystals in Ref. \cite{Liu2017} was discerned only by neutron diffraction measurements. And more importantly, that the saturated moment at 300 K and at 5 K are practically the same indicating a FM component that develops at higher temperatures than $T_{\rm N}$. This clearly suggests the FM in that crystal cannot arise from canted moments in the C-type AFM phase, but could arise from a FM impurity phase that orders well above $T_{\rm N}$. Below $T_{\rm N}$, magnetization of Batch A is consistent with ordered AFM and axial anisotropy.  At higher fields, dHvA oscillations originating from 2D charge carriers are evident for \textbf{H}$||$\textit{\textbf{a}} as discussed below. The magnetization of Batch B, including anisotropy and dHvA are nearly identical to batch A.  However, a saturated FM component is clearly visible in Batch B.

\begin{figure}
\includegraphics[width=1.0\linewidth]{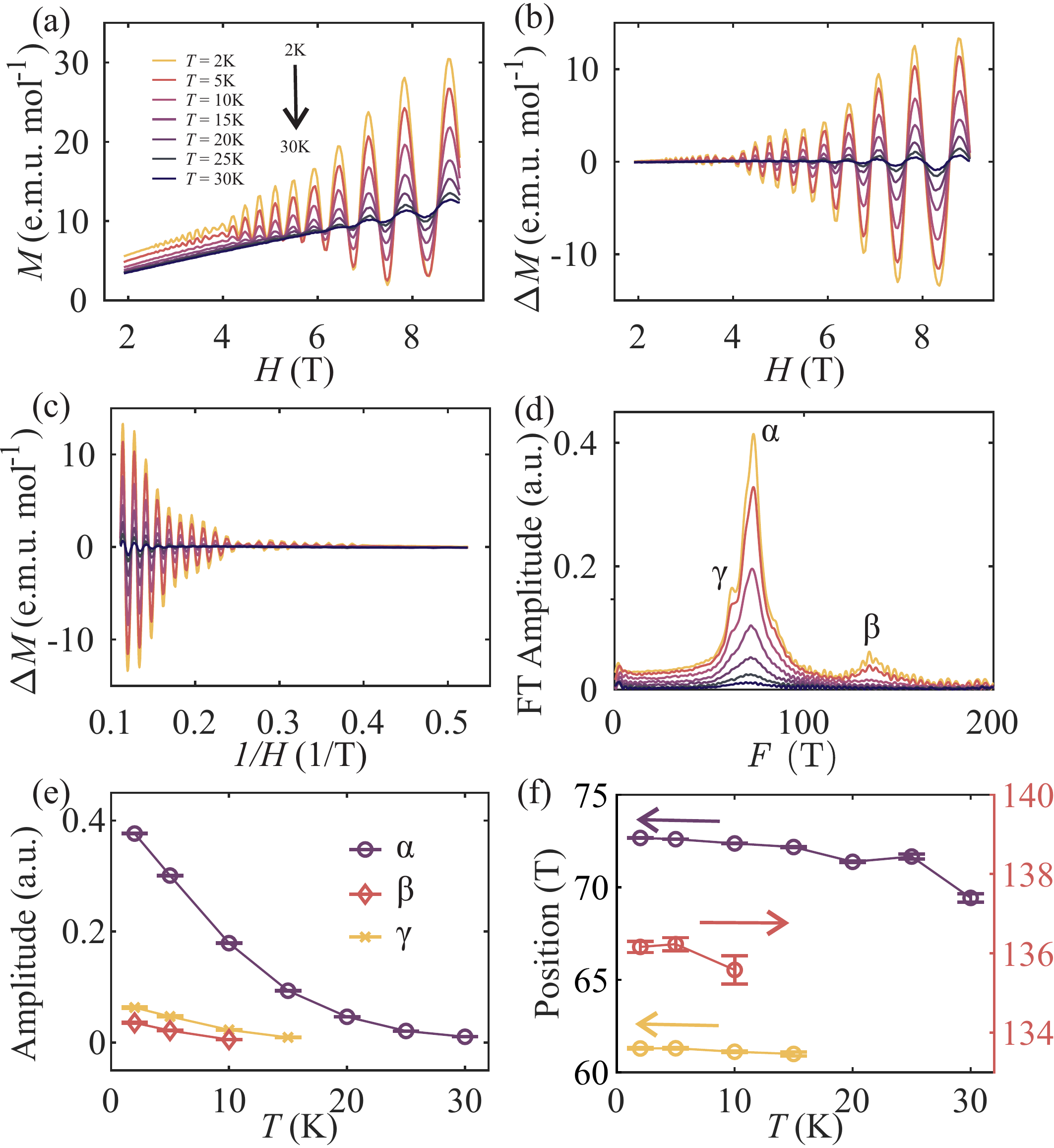}
\caption{(Color online) (a) Magnetization versus applied magnetic field perpendicular to the MnSb planes of SrMnSb$_2$ single crystal (Batch A) at various temperatures as indicated showing de Haas-van Alphen (dHvA) oscillations. (b) Oscillatory component of the magnetization $\Delta M$ obtained by subtracting an analytical function $f(B)=c_1{\rm erf}(c_2H)+c_3H+c_4$ from data shown in (a) with adjusted parameters $c_i$. (c) Same data as in (b) plotted versus $1/H$.  (d) Numerical FT spectra obtained from data shown in (c). (e) Integrated intensities versus temperature of the various FT peaks as indicated in (d). (f) Peak position of the FT for the 3 peaks identified in (d). }
\label{Fig:DHVA}
\end{figure}

In Fig. \ref{Fig:DHVA} we provide detailed analysis of the dHvA oscillations of Batch A crystals. The $M (H)$ exhibits strong oscillations that are temperature dependent as shown in Fig. \ref{Fig:DHVA}(a).  To obtain the Fourier transform (FT) spectra from the oscillating  $M (H)$, we first subtract from the data in (a)  an analytical function $f(H)=c_1{\rm erf}(c_2H)+c_3H+c_4$ with adjustable parameters $c_i$ to obtain the oscillatory portion as shown in Fig. \ref{Fig:DHVA}(b). We then create a plot of the same data as a function of $1/H$ in Fig. \ref{Fig:DHVA}(c).

The FT spectra shown in Fig. \ref{Fig:DHVA}(d) display a prominent peak ($\alpha$) at 73.1 T which is slightly higher than the main peak in FT spectra (66.7 T) for Type C crystals in Ref. \cite{Liu2017}. There exist a shoulder ($\gamma$) at 62.1 T beside the main peak and the third peak ($\beta$) at 134.0 T in the FT spectra. These two peaks were also reported in Ref. \cite{Liu2017}, but at different location. The main FT amplitude is temperature dependent and seems to vanish at $\sim 30$ K.
The FT peaks in Fig. \ref{Fig:DHVA}(d) are modeled as a sum of Lorntzian-shaped peaks which are labeled by $\alpha$, $\beta$ and $\gamma$. Figures \ref{Fig:DHVA} (e) and (f) show the temperature dependence of the peak intensities and positions as a function of temperature. The temperature dependence of the main peak at $H_{FT} = 73.1$ T can be modeled with the Lifshitz-Kosevich equation to obtain the charge carrier's effective mass ($m^*$) as given in Ref. \cite{Kartsovnik2004}
\begin{equation}
P_{FT}= C\frac{K Tm^*/H}{\sinh{(K Tm^*/H)}}
\end{equation}
where $C$ is a scale factor, $K = 2\pi^2k_B m_e/(\hbar e) \eqsim 14.69$ T/K. $H$ corresponds to the inverse average of the used field window which the Fourier analysis is performed, such that $H =[(1/ H_{\rm start}+1/ H_{\rm end})/2]^{-1} = 7.2$ T with $H_{\rm start}=6$ and $H_{\rm end}=9$ T. This is the same method used in Ref. \cite{Liu2017}. The fit to the data in Fig. \ref{Fig:DHVA}(e) yields an effective mass of $m^*=0.06(1)$, which is consistent with the results in Ref. \cite{Liu2017}.

\begin{figure}
\centering
\includegraphics[width=1.0\linewidth]{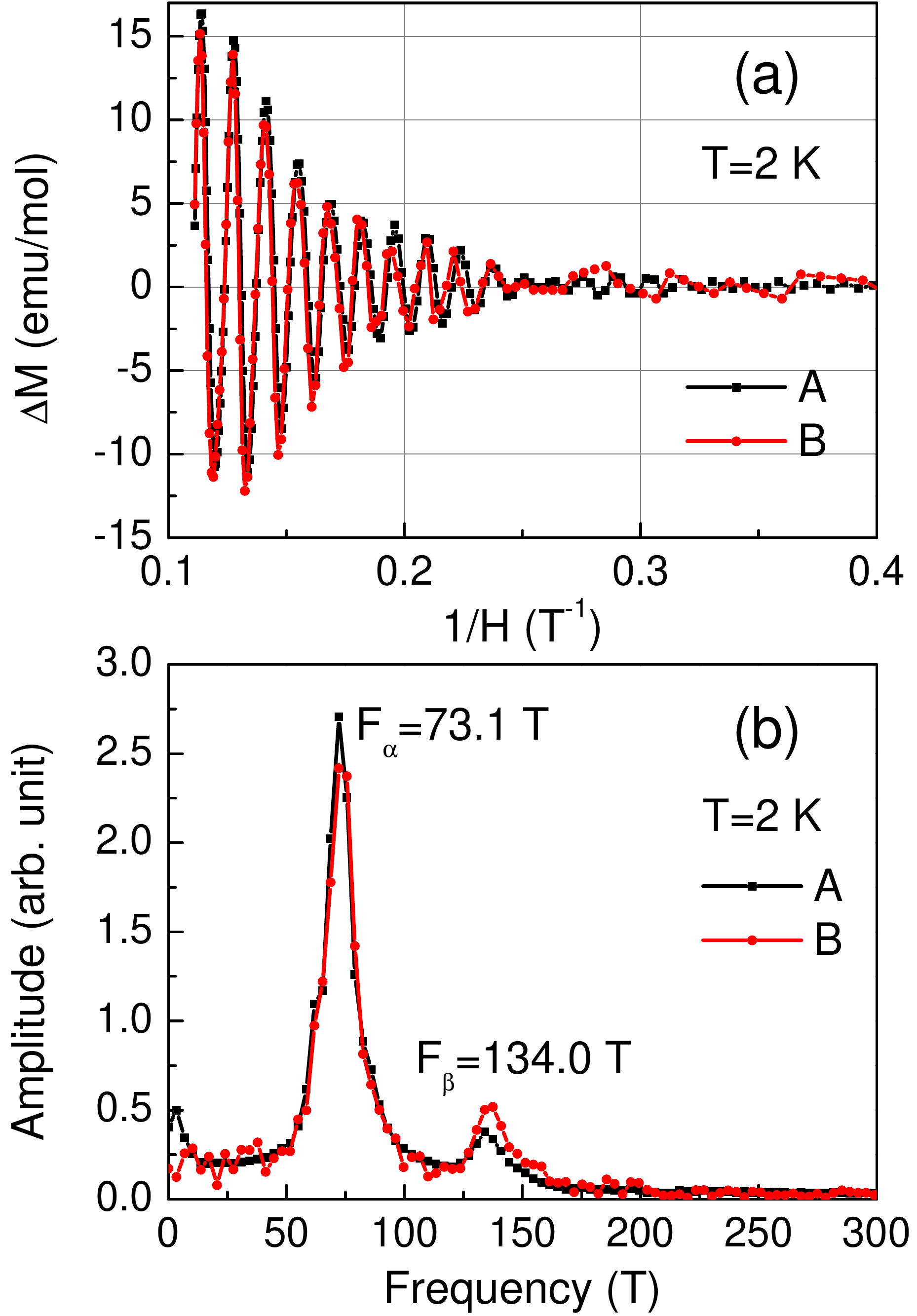}
\caption{(Color online) (a) Oscillatory component of magnetization at 2 K as a function of inverse field for Batches A and B SrMnSb$_2$ crystals. (b) The spectra of Fast Fourier transformation for the data in (a).} 
\label{FT:Type12}
\end{figure}

Figure \ref{FT:Type12}(a) shows a comparison of dHvA oscillations of Batches A and B crystals at 2 K.  The fact that the fundamental frequency for both samples is the same is of paramount importance regarding the stoichiometry of the two samples.  We emphasize that in the event of non-stoichiometric chemical formula for any of the two types of crystals (for instance deficiency of Sr or Mn) should result in doping with electrons or holes. Any resulting changes in the chemical potential are expected to significantly modify the Fermi surface (FS) and subsequently the FT frequency. Thus, the fact that both Batch A and B crystals show the same FT spectra is indirect evidence that they both have the same FS. 

\begin{figure}
\centering
\includegraphics[width=0.72\linewidth]{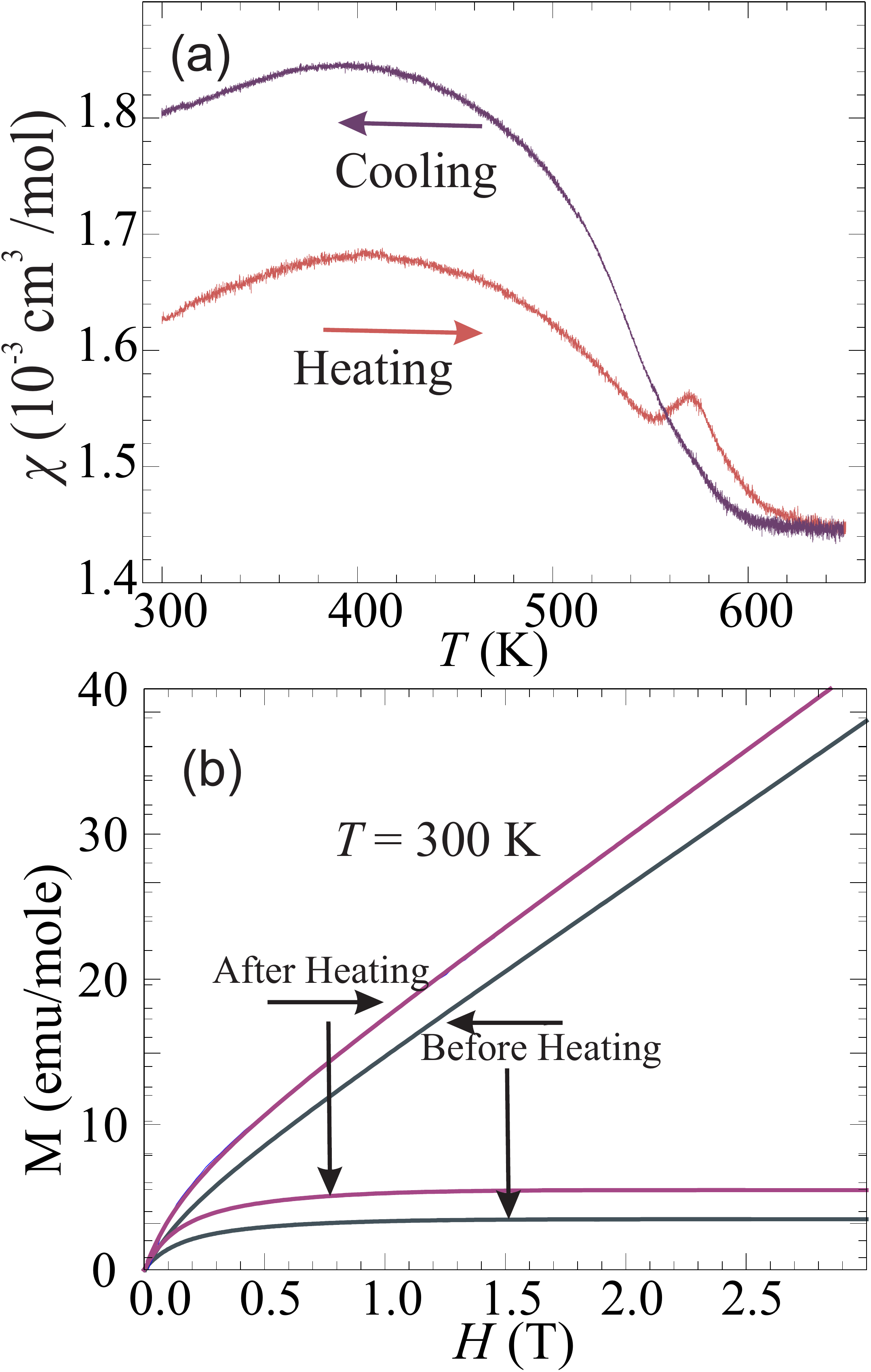}
\caption{(Color online) (a) Temperature dependence of magnetic susceptibility $\chi(T)$ in \textbf{H}$\perp$\textit{\textbf{a}} axis for a Batch A crystal cycled from 300 K to 650 K and back to 300 K at $H=1$ T. Upon heating a broad peak emerges at $T \sim 420$ K, a typical feature of two-dimensional (2D) magnetic systems. Further increase in  temperature gives rise to a strong peak  at $T \sim 570$ K. Upon cooling the susceptibility curve behaves like a ferromagnet with $T_C \sim 580$ K. (b) $M(H)$ data at $T=300$ K for the same crystal as that measured in (a), before and after heating. The lower curves show the corresponding extracted FM components. } 
\label{Fig:Susce3}
\end{figure}

Clues to the origin of the FM signal observed in Batch B samples can be obtained from high temperature measurements of the magnetic susceptibility. Figure \ref{Fig:Susce3}(a) shows $\chi (T) $ of Batch A crystals measured in the temperature ($T$) range of 300 to 650 K for \textbf{H}$\perp$\textit{\textbf{a}} axis. The susceptibility data over $T=2$ to 340 K are the same as those shown in Fig. \ref{Fig:Mag-I}(a). They exhibit weak temperature dependence as expected for an AFM with ordered moments oriented perpendicular to the applied magnetic field.  Above $T_{\rm N}$ the susceptibility increases slightly with temperature up to about 420 K, where a broad peak is about to form. Thus, 2D AFM correlations develop at temperatures that are much higher than the 3D N{\'e}el temperature $T_{\rm N}$, and  consequently, $\chi(T)$  shows a very broad peak at temperatures that are much higher than $T_{\rm N}$ at the onset of emerging 2D magnetic correlations \cite{Curely1998}. This kind of behavior is typical of AFM systems with in-plane exchange coupling among nearest neighbors (NN; $J_{\rm 2D}$) that is much stronger than the inter-plane coupling $J_{\rm a}$,  as has been observed in similar layered systems \cite{Vaknin1989,Sangeetha2016}. This behavior, is common to various compounds that posses MnSb layers such as AMnSb$_2$  and $R$Mn$Pn$O ($R =$ La, Ce, Pr ...; $Pn =$ P, As, Sb compounds) that all settle into a G- or C-type AFM ground state \cite{An2009,Singh2009,Singh2009b,Zhang2015,McGuire2016,Zhang2016,Liu2018}. However, at $T \sim 570$ K a peak emerges that disappears above $T \sim 600$ K. Upon cooling the same sample, the susceptibility displays a characteristic FM transition with a $T_C$$\simeq 580$ K. This irreversible behavior is reminiscent of the behavior of Batch B samples and clear evidence that single crystals are highly reactive and that a FM impurity is being formed. Figure \ref{Fig:Susce3}(b) shows magnetization versus $H$ that for the same crystal before and after heating it to 650 K.  We model the magnetization as follows: $M(H) =A{\rm erf}(\sigma H) + BH+ C$, where erf($\sigma H$) is the Error function, which we associate with the FM contribution. As one can see in Fig. \ref{Fig:Susce3}(b), the FM contribution, i.e., the $A{\rm erf}(\sigma H)$ component, indeed increases after heating the crystal. We note that since this is a degradation process of the crystals, it is reasonable to expect variations of FM component as observed for different crystals. In fact, visual inspection of the unloaded crystal shows grayish powder on the surface that before these measurements had shiny black surfaces. To obtain more insight on the nature of this surface transformation, we performed powder neutron diffraction of crushed crystals into powder at elevated temperatures, discussed below.

\subsection{Neutron diffraction} 

\begin{figure}
\includegraphics[width=2.6 in]{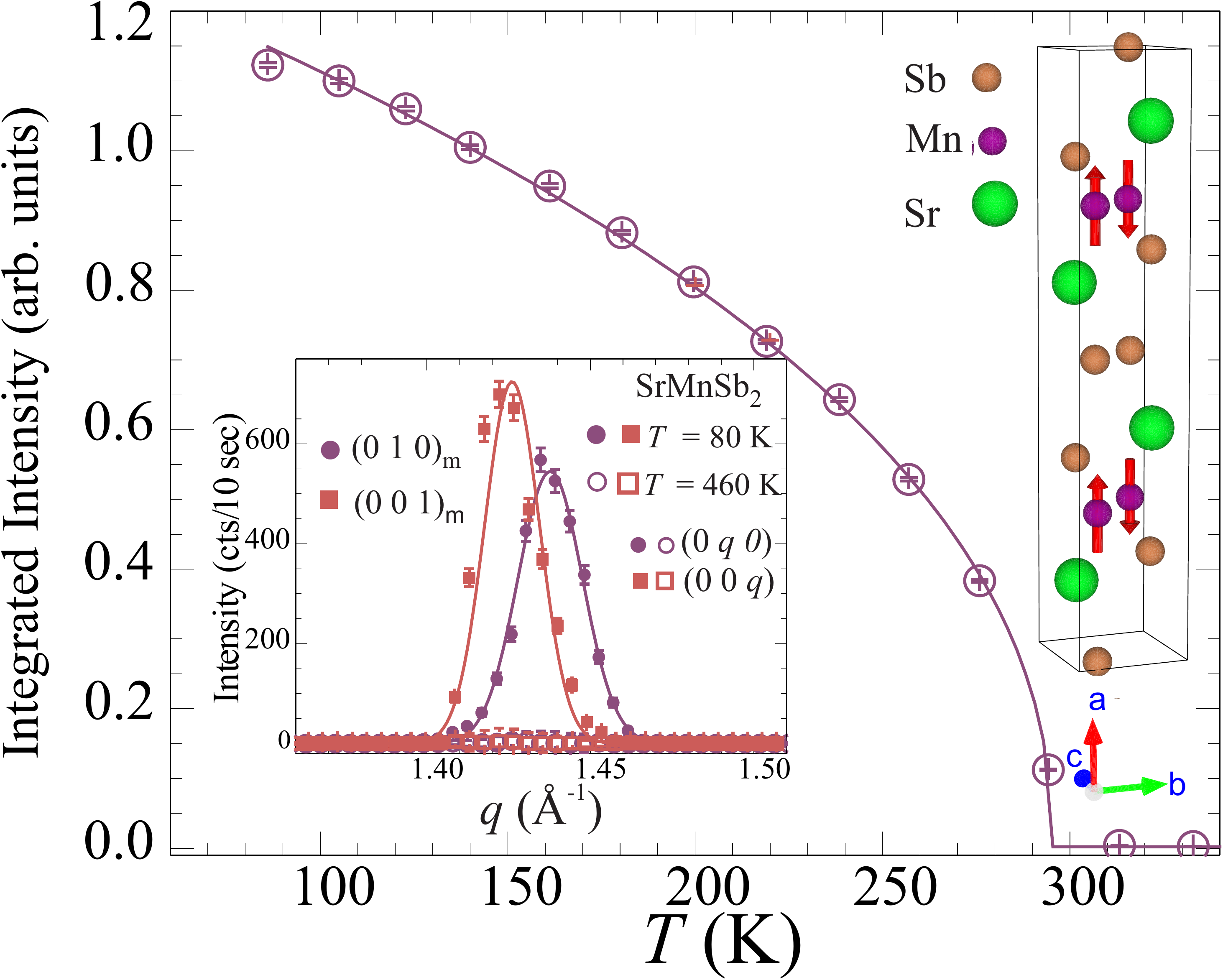}
\caption{(Color online) Intensity versus temperature of the magnetic (010) including a fit to a power law $I \propto (1-T/T_N)^{2\beta}$ (solid line) that yields $T_N = 297(2)$  K and $\beta = 0.227(2)$. Right inset illustrates atomic chemical and magnetic structure and left Inset shows intensity of magnetic Bragg reflections (0,1,0) and (0,0,1) versus momentum transfer for single crystal SrMnSb$_2$ below and above $T_{\rm N}$. As is evident, the crystal is to a large extent not twinned. }
\label{Fig:OP}
\end{figure}

Figure \ref{Fig:OP} shows the intensities of the (0,1,0) and (0,0,1) AFM Bragg peaks at 80 and 460 K for a Batch A SrMnSb$_2$ crystal. The distinct (un-split) planar orthorhombic peaks indicate a single crystal with almost no detectable orthorhombic twinning. These and other measured magnetic peaks are consistent with the previously reported C-type AFM structure \cite{Liu2017}. The intensity versus temperature of the (010) magnetic Bragg reflection fits (solid line) to a power law $I \propto (1-T/T_N)^{2\beta}$ with $T_N = 297(2)$  K and $\beta = 0.227(2)$. The inset in (c) illustrates the chemical and magnetic structures.  Using a few magnetic and nuclear Bragg reflections, we confirm the C-type AFM order and calculate an average staggered magnetic moment per Mn$^{2+}$ of 4.0(3) $\mu_{\textrm{B}}$ at 80 K. 

\begin{figure}
\includegraphics[width=3. in]{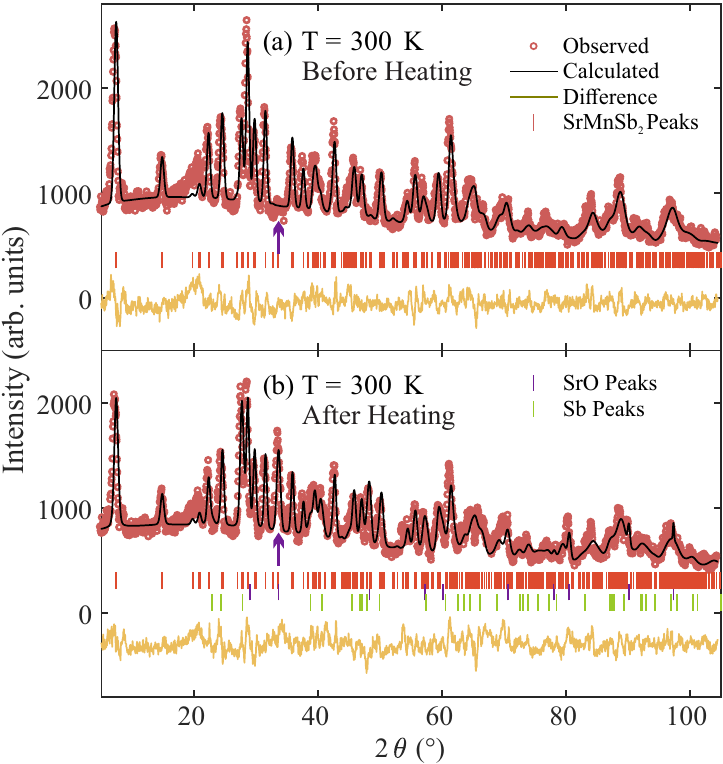}
\caption{(Color online)  (a) Neutron powder diffraction (conducted at the MURR) obtained at 300 K under $10^{-6}$ Torr, and Rietveld refinement show the sample is free of secondary phases.  (b) Similar to (a), but after heating the sample to 600 K. The structural refinement required the addition of SrO and Sb metal as additional minority phases.  This suggests the formation of small  but undetectable amounts of MnSb which is FM in the temperature range 560 to 600 K.}
\label{Fig:diffraction}
\end{figure}

To further clarify the issue of the emergence of a FM impurity, we have conducted neutron powder diffraction on crushed single crystals to high temperatures to identify  minority phases that may be associated with the FM component found in $\chi(T)$. Figure \ref{Fig:diffraction}(a) shows neutron diffraction patterns at 300 K of the as-prepared sample, along with a structural refinement that indicates the sample is free of minority phases. However, after heating the sample to 600 K under vacuum ($10^{-6}$ Torr) and cooling it back to 300 K, the refinement of the pattern in Fig. \ref{Fig:diffraction}(b) indicates the formation of SrO and smaller amounts of Sb metal. According to our high temperature susceptibility measurements (see Fig.\ \ref{Fig:Susce3}), the oxidation of the surface is most pronounced above 500 K.

\begin{figure}
\includegraphics[width=3.2 in]{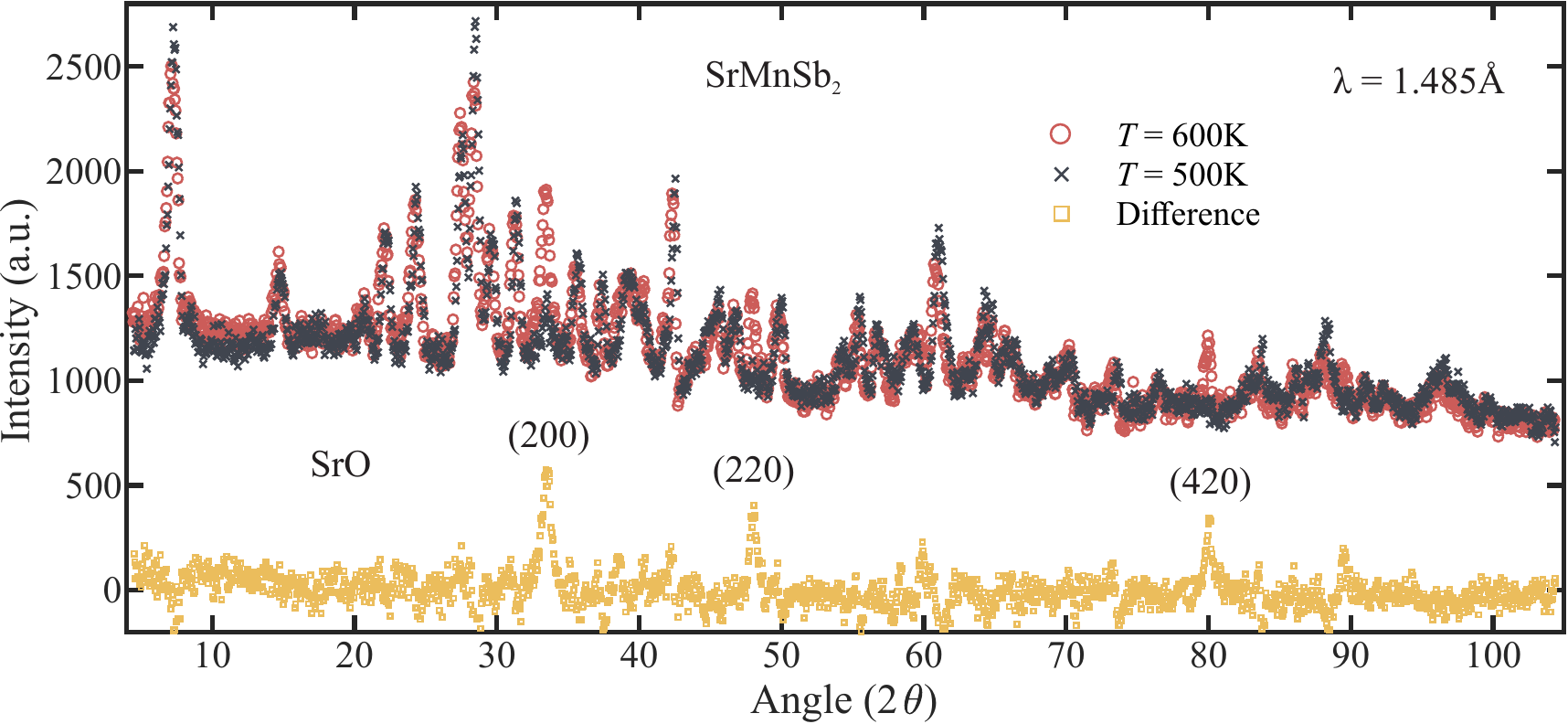}
\caption{(Color online)  Neutron diffraction patterns from polycrystalline SrMnSb$_2$ sample at 500 and 600 K and the difference between them showing the emergence of SrO as an impurity phase.  The labeled SrO peaks are the most intense based on calculations of powder diffraction of SrO.  The measurements were conducted under vacuum at an average pressure of $10^{-6}$ Torr.}
\label{Fig:Diff}
\end{figure}

Figure \ref{Fig:Diff} shows neutron diffraction patterns taken at 500 and 600 K and the difference between them. We chose to subtract the 500 K from the 600 K data as they are close in temperature with minimal shifts of peak-positions due to thermal expansion.  The difference between the two patterns indicates prominent peaks that we can readily identify as SrO. This is consistent with our diffraction analysis of the same powder after the sample temperature is lowered from 600 K back to 300 K. This shows that the surfaces of SrMnSb$_2$ are highly reactive to oxygen even in minute amounts, especially at temperatures higher than 500 K.

Based on these observations, we propose that the FM signal is due to the presence of minute amounts of MnSb which depends on composition and on thermal annealing procedures undergoes a FM transition between $530 \leq T \leq 600 $ K \cite{Takei1963}. Assuming the saturated FM moment shown in Fig. \ref{Fig:Susce3} and the reported average ordered FM moment of MnSb ($\sim$3 $\mu_{\textrm{B}}$) we estimate the amount of MnSb in the sample by weight as $\sim 0.2$\%. Such an amount ($\sim 3$ mg) in our neutron powder diffraction measurements is too small to be observed in the diffraction pattern. Table\ \ref{tab:NPD} lists the structural parameters obtained from the neutron diffraction measurements at the various measured temperatures. We find no evidence that a structural transition occurs from 300 to 600 K. 

\begin{table*}[ht]
\caption{\label{Tab:FitParams} Fit parameters obtained from Rietveld refinements of the powder neutron diffraction patterns at various temperatures using the orthorhombic $Pnma$ space group. $a$, $b$, $c$, and $V$ are the unit cell parameters and volume, respectively. $z_{\rm Sb}$ represents the Sb $z$ position in the crystal structure. $\chi^2$ gives the overall value of the goodness of fit. The error (one standard deviation) is in the last digit of a quantity and is shown in parentheses.}
\label{tab:NPD}
\centering
\begin{ruledtabular}
\begin{tabular}{ccccccc}
$T $ & $a$ & $b$ & $c$ & $V$ &  other phases&  $\chi^2$ \\
(K) & ($\textrm{\AA}$) & ($\textrm{\AA}$) &($\textrm{\AA}$) 	& ($\textrm{\AA}^3$) &    & \\
\hline
300 (before) & 23.044(5)& 4.380(1) & 4.423(1) & 446.5(1)&  none & 4.84 \\
400 & 23.093(5) & 4.389(2) & 4.431(2) & 449.0(2) & none & 5.12 \\
500 & 23.143(6) & 4.395(5) &4.437(2) & 451.3(5) &  none & 5.59 \\
600 &23.182(7)& 4.401(2) &4.442(4)& 453.2(4) & SrO, Sb & 5.12 \\
300 (after) & 23.034(6) & 4.383(2) & 4.420(2)& 446.2(2) &  SrO, Sb & 6.33 \\

\end{tabular}
\end{ruledtabular}
\end{table*}

\subsection{X-ray magnetic circular dichroism - XMCD}

\begin{figure}
\centering
\includegraphics[width=1.0\linewidth]{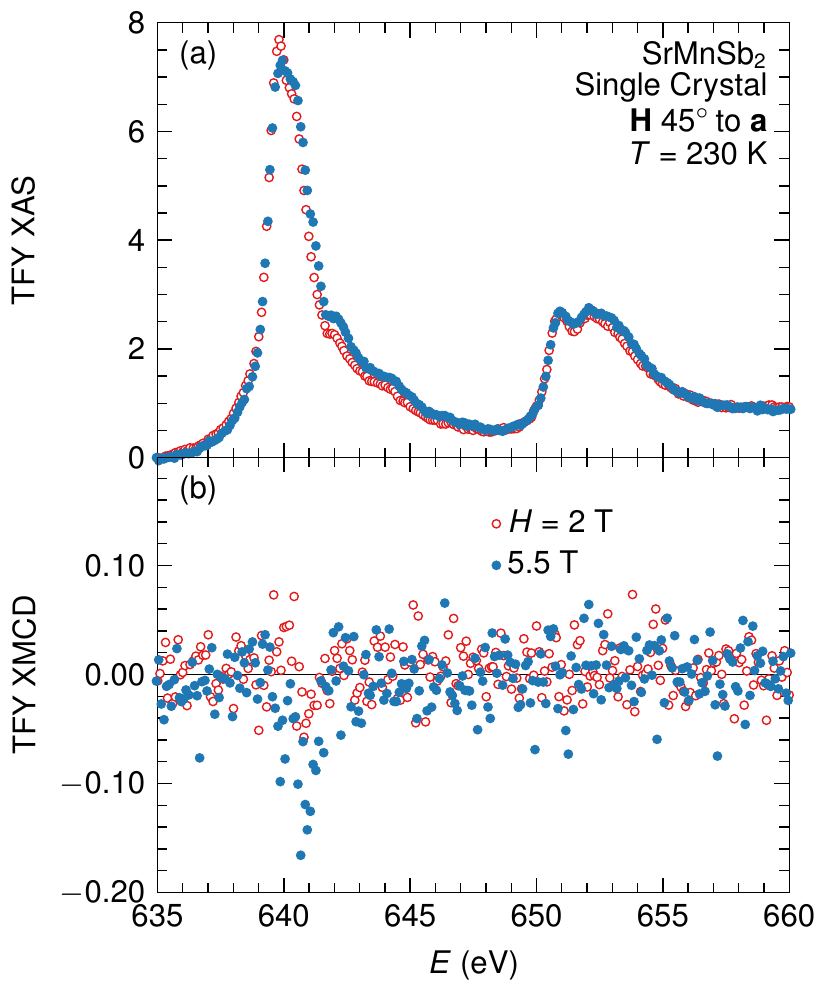}
\caption{Normalized XAS (a) and XMCD (b) total-fluorescence-yield (TFY) signals for the Mn $L_{3}$ and  $L_{2}$ edges for a single-crystal sample of SrMnSb$_{2}$ from Batch A.  The XAS signal has been normalized to 1 at energies well above the edge, and the XMCD data is normalized to the XAS data.  Data are for $H=2$ and 5.5 T and $T=230$~K, and are not corrected for self-absorption effects.  The single-crystal sample was oriented with its \textbf{a} axis $45 \degree$~away from the incoming beam.}
\label{XMCD_TFY}
\end{figure}

The reported canted C-type AFM order in SrMnSb$_{2}$ is associated with the Mn cations canted away from the \textit{\textbf{a}} axis such that a ferromagnetically ordered moment of $\approx0.2~\mu_{\textrm{B}}/$Mn is oriented within the \textit{\textbf{bc}} plane \cite{Liu2017}. XMCD is a sensitive probe that can detect small amounts of FM component with element specificity. The Mn $L_{3}$ and $L_{2}$ X-ray absorption edges measured in this experiment correspond to transitions from the $2p_{\frac{3}{2}}$ to $3d$ and $2p_{\frac{1}{2}}$ to $3d$ electronic states, respectively.   Thus, magnetization arising from the canted-AFM order of the Mn with that magnitude should be detectable by our XMCD measurements. We also note that the total-electron-yield (TEY) measurements (data not shown) only probe $\sim5$~nm into the sample, whereas total-fluorescence yield (TFY)  probe $\sim50$--$100$~nm beneath the sample's surface \cite{Lang2003}.  This means TFY measurements are more reliable in probing the intrinsic properties of the sample compared to TEY being more surface sensitive.  Figure \ref{XMCD_TFY} shows XAS and XMCD spectra at the Mn $L_{3}$ and $L_{2}$ absorption edges in the TFY mode of Batch A crystal at $T=230$~K and $H=2$ and $5.5$~T.  The XAS data are normalized to zero below the $L_{3}$ edge and to one above the $L_{2}$ edge, and the XMCD data are normalized to the resulting step over both edges.  A very weak signal in the XMCD data at the $L_{3}$ edge at $H=2$ T slightly increases at  $5.5$~T.  Such field dependent response is more characteristic of field induced canting of an AFM system as that shown in Fig. \ref{Fig:Mag-I}(c) for \textbf{H}$\perp$\textit{\textbf{a}} rather than a spontaneous FM component that saturates at much lower fields (see Fig. \ref{Fig:Mag-II}). 

Applying of standard  sum-rule analysis to the spectra \cite{Lang2003} at $T=235$ K and $H=5.5$ T, and assuming electronic configuration of  Mn$^{2+}$ yields  $m_{\textrm{\textbf{bc}}}=m_{\textrm{orb}}+m_{\textrm{s}}\leq 0.01  \mu_{\textrm{B}}/$Mn, where $m_{\textrm{orb}}$ and $m_{\textrm{s}}$ are the orbital and spin projection of the total magnetic moment in the \textit{\textbf{bc}} plane.  This value is consistent with the susceptibility measurements for Batch A crystals in this study and Type C crystals in Ref. \cite{Liu2017}. 
\begin{figure}
\centering
\includegraphics[width=3.4 in]{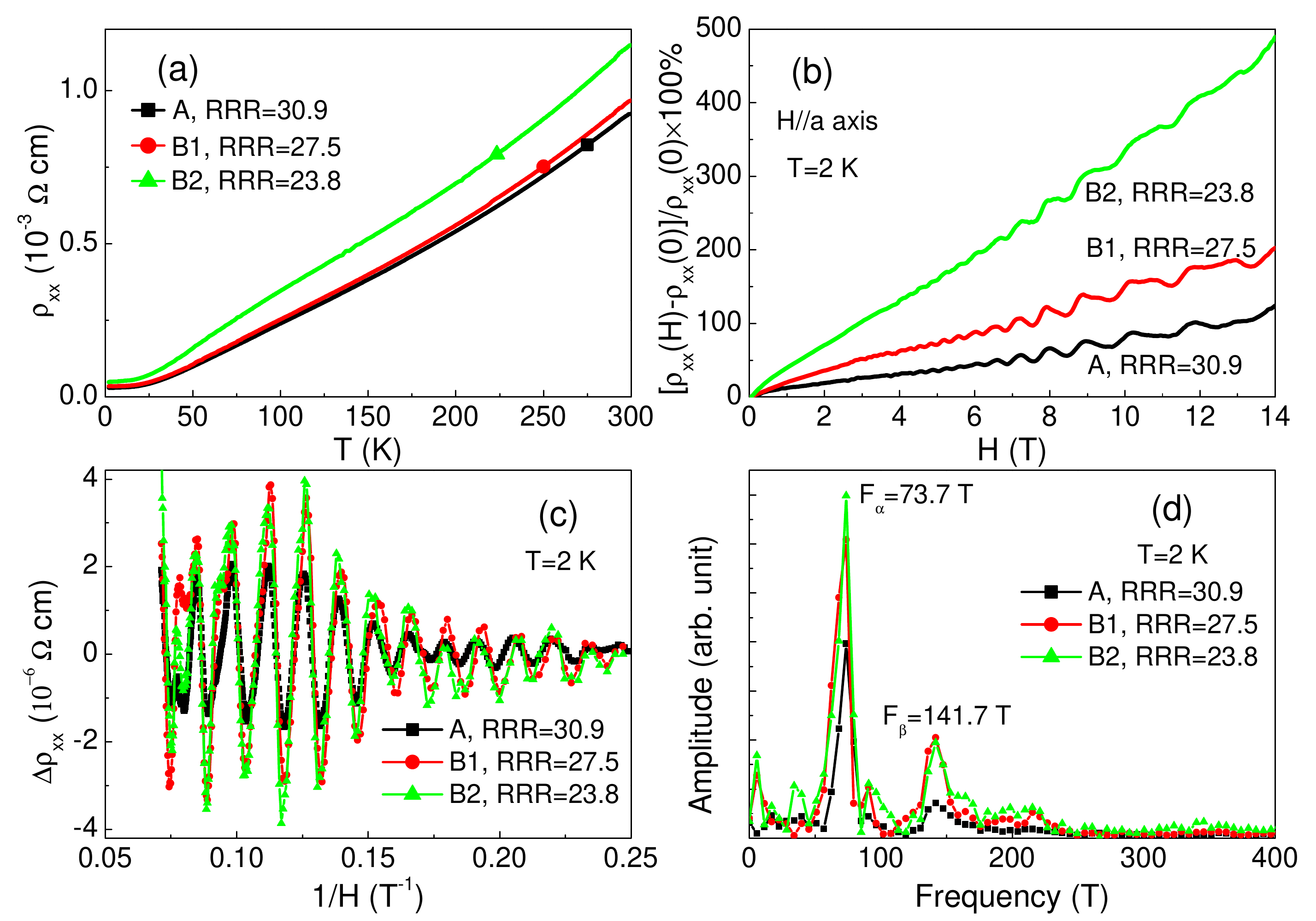}
\caption{(Color online) (a) Temperature dependence of resistivity $\rho_{xx}$ for SrMnSb$_2$ crystals from Batch A and Batch B (B1 and B2) with different $RRR$ values at zero magnetic field. (b) The magnetoresistance with \textbf{H}$||$\textit{\textbf{a}} configuration showing a clear Shubnikov-de Haas (SdH) oscillations up to 14 T at $T=2$ K. (c) Oscillatory component of resistivity at 2 K as a function of inverse magnetic field $1/H$ for the three crystals as indicated. (d) Fast Fourier transformation (FFT) spectra of oscillation data in (c).}
\label{Fig:SdH}
\end{figure}

\subsection{Shubnikov-de Haas effect and Hall coefficient}
WSMs can be described as `bad metals' \cite{Hosur2013}.  Figure \ref{Fig:SdH}(a) shows the temperature dependence of resistivity $\rho_{xx}$ with  residual resistivity ratio $RRR\equiv R(300 K)/R(2 K)$ values of 30.9, 27.5 and 23.8, as indicated. To ascertain whether the FM component has any effect on the transport properties we examine the transport properties of the two batches of crystals under magnetic field. Figure \ref{Fig:SdH}(b) shows magneto-resistatnce $MR=[\rho_{xx}(H)-\rho_{xx}(0)]/(\rho_{xx}(0))\times100$\% at $T=2$ K with \textbf{H}$||$\textit{\textbf{a}}. Clear Shubnikov-de Haas effect (SdH) oscillations occur up to 14 T. Note that the magnetoresistance ($MR$) is most pronounced in crystals with a low $RRR$ value and large saturated magnetization.

Figure \ref{Fig:SdH}(c) shows the SdH oscillatory component as a function of inverse field at $T=2$ K after subtracting a baseline. The amplitude of oscillations of the Batch B crystal is larger than that of the Batch A crystals. The FFT analysis for the three samples give primary SdH frequencies of $F_{\alpha}=73.7(5)$ T and a minor one at $F_{\beta}=141.7(8)$ T, as shown in Fig. \ref{Fig:SdH}(d). These values are consistent within error with dHvA oscillations obtained in the magnetization measurements above regardless of batch or $RRR$ value. Figure \ref{Fig:SdH}(c) shows that the amplitude of the FFT spectra is highest for the sample with lowest $RRR$ value. However, the two frequencies $F_{\alpha}$ and $F_{\beta}$ in Fig. \ref{Fig:SdH}(d) are practically the same, indicating that the size of the saturated magnetization does not modify the FS. 

\begin{figure}
\includegraphics[width=3.4 in]{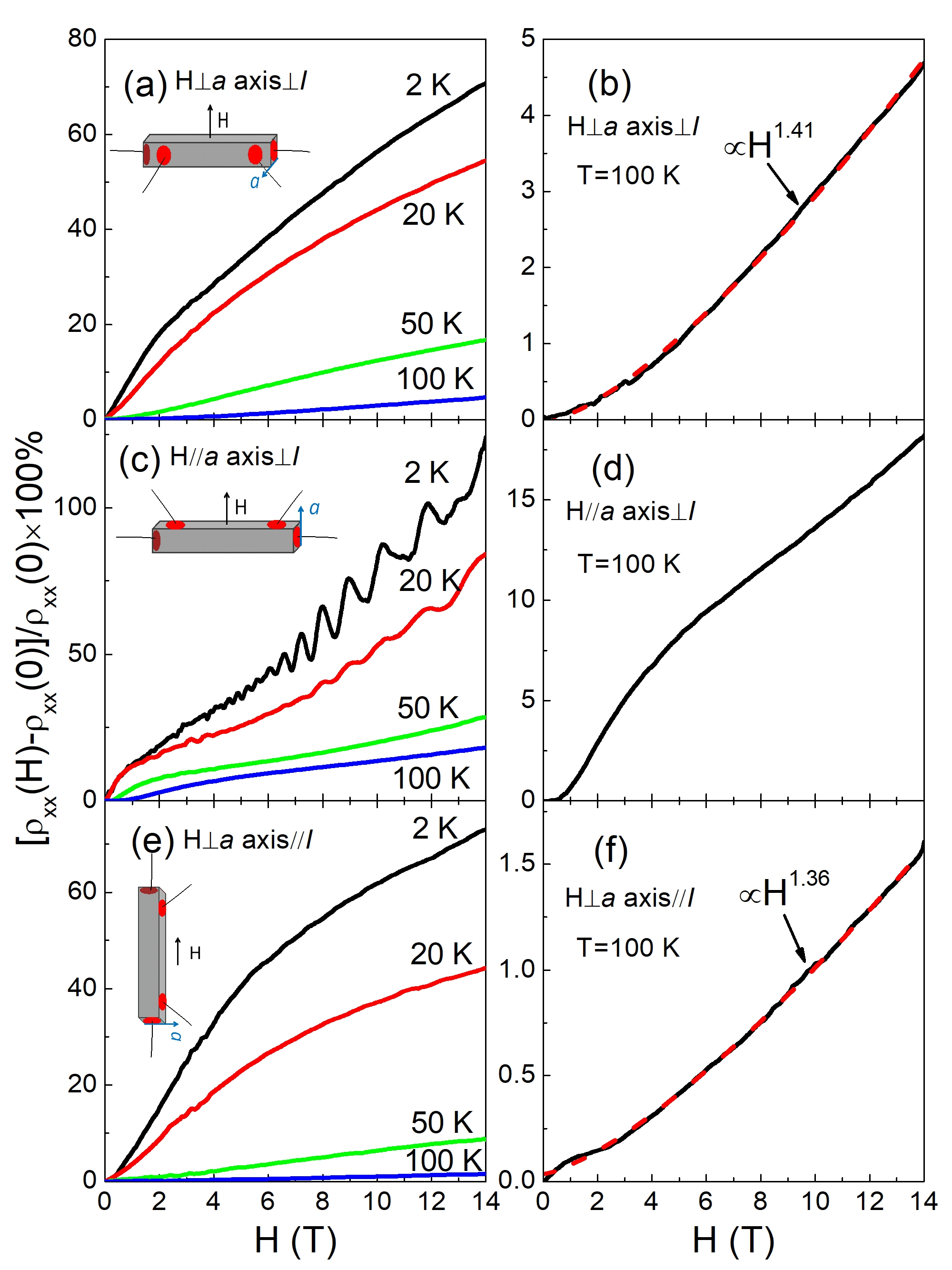}
\caption{(Color online)  (Color online) The magnetoresistance (MR) of the Batch A crystals with $RRR=30.9$ in the configurations: (a) \textbf{H}$\perp$\textit{\textbf{a}}$\perp$\textit{\textbf{I}}; (c) \textbf{H}$||$\textit{\textbf{a}}$\perp$\textit{\textbf{I}}; (e) \textbf{H}$\perp$\textit{\textbf{a}}$||$\textit{\textbf{I}} at $T=2$, 20, 50 and 100 K and the MR at for the corresponding configurations in (b), (d), and (f) at $T=100$ K. The MR curves in (b) \textbf{H}$\perp$\textit{\textbf{a}}$\perp$\textit{\textbf{I}}; (f)  \textbf{H}$\perp$\textit{\textbf{a}}$||$\textit{\textbf{I}} obey a power law behavior. Configuration (c) shows a different behavior at 100 K, but exhibits SdH oscillations at low temperatures.}
\label{Fig:MR}
\end{figure}

Figure \ref{Fig:MR} shows the temperature dependence of the longitudinal resistivity $\rho_{xx}$ and MR effect in SrMnSb$_2$ crystals from Batch A under various magnetic fields and driven current configurations: (a), (b) \textbf{H}$\perp$\textit{\textbf{a}}$\perp$\textit{\textbf{I}};  (c), (d)  \textbf{H}$||$\textit{\textbf{a}}$\perp$\textit{\textbf{I}}; (e), (f) \textbf{H}$\perp$\textit{\textbf{a}}$||$\textit{\textbf{I}}. This is the same sample as the one with $RRR=30.9$ and measured in Fig. \ref{Fig:SdH}. The most relevant feature in the MR measurements is the observation of SdH effect only in the case of \textbf{H}$||$\textit{\textbf{a}}$\perp$\textit{\textbf{I}}, as shown in Fig. \ref{Fig:MR}(c). Furthermore, the MR  is most pronounced for \textbf{H}$||$\textit{\textbf{a}}$\perp$\textit{\textbf{I}} while it becomes smallest for \textbf{H}$\perp$\textit{\textbf{a}}$||$\textit{\textbf{I}}. For the configuration with \textbf{H}$||$\textit{\textbf{a}}$\perp$\textit{\textbf{I}} [Figure \ref{Fig:MR}(c)], the MR roughly follows a linear field dependence and starts to oscillate above $H=4$ T at $T=2$ K. With increasing temperature to 100 K, the MR still roughly follows a linear field dependence at high field region but with a slightly downward bending, as shown in Fig. \ref{Fig:MR}(d). In both \textbf{H}$\perp$\textit{\textbf{a}}$\perp$\textit{\textbf{I}} and \textbf{H}$\perp$\textit{\textbf{a}}$||$\textit{\textbf{I}} configurations, the MR bends down (i.e. tends to saturate) with increasing fields at low temperatures.  However, at $T=100$ K, the MR curves can be fitted to a power law $MR \propto H^n$ within the field range  $0\leq H \leq 14$ T, with $n =1.41$ and 1.36 for the configurations shown in Figs. \ref{Fig:MR}(b) and (f).

It should be pointed out that the conductivity in a WSM is predicted to be proportional to the magnetic field or, equivalently, the resistivity is expected to decrease with increasing magnetic field, i.e., showing a negative MR effect \cite{Hosur2013}. We have measured Batch-A and -B SrMnSb$_2$ crystals in different configurations and did not find evidence for a negative MR effect.

To summarize this part, the FFT analysis of dHvA and SdH oscillations shows that the dominant frequency occurs at $F_\alpha \simeq 73$ T within error for SrMnSb$_2$ single crystals from both Batches despite their different FM signal and $RRR$ values. This is a very important result, as this main Fourier frequency is sensitive to the FS topology and should be strongly dependent on doping. Batch A and B samples show different FM signals but have nearly identical FS and effective masses.  

\begin{figure}
\includegraphics[width=3.4 in]{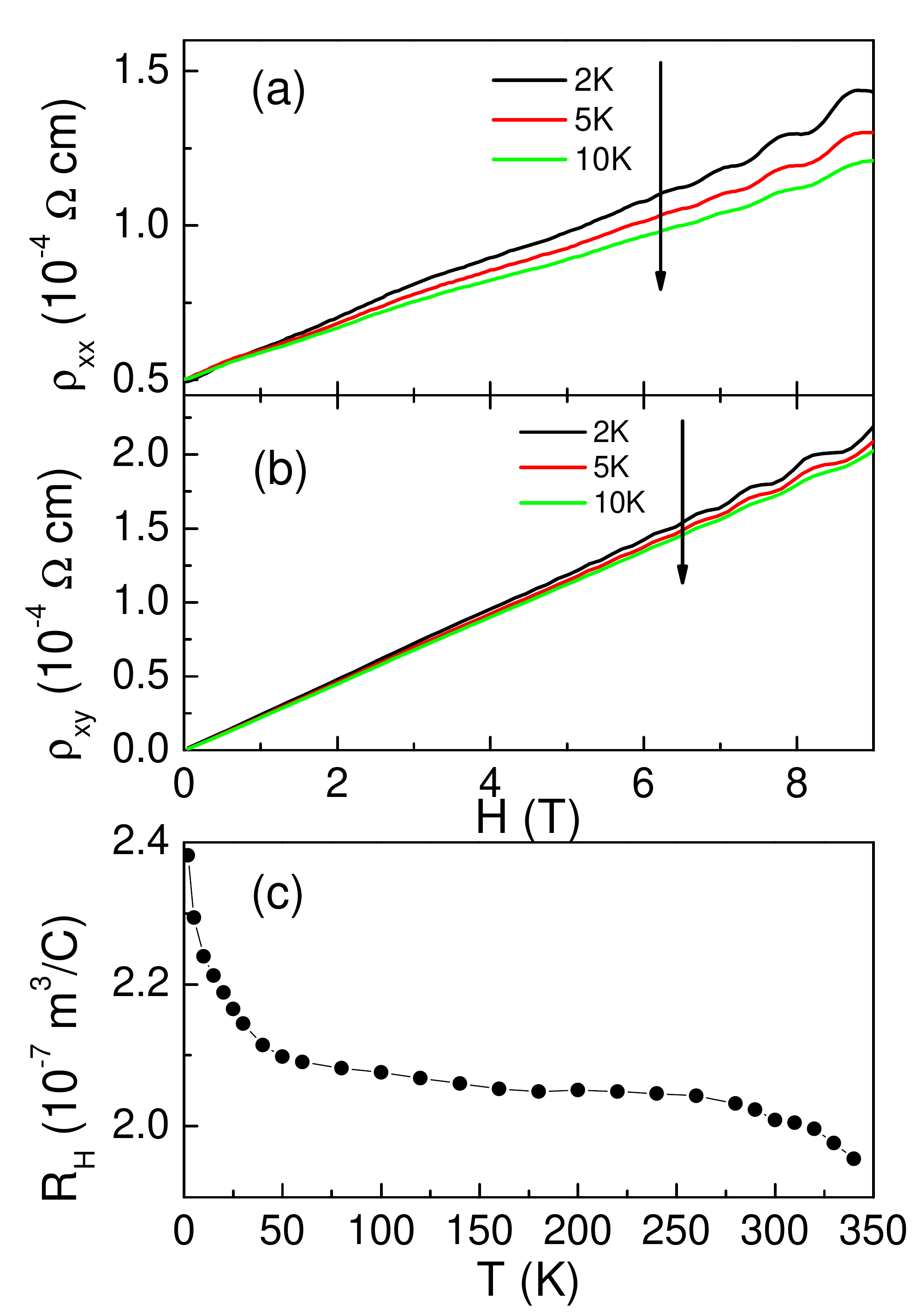}
\caption{(Color online) (a) The magnetic field dependence of longitudinal resistivity $\rho_{xx}$ at $T=2$, 5, and 10 K for a Batch A crystal. Arrow indicates the increase of measured temperatures. (b) The magnetic field dependence of transverse (Hall) resistivity $\rho_{xy}$ at $T=2$, 5, and 10 K. (c) Temperature dependence of Hall coefficient $R_{\rm H}$. A small kink is in coincidence with the AFM transition temperature observed in susceptibility measurements. }
\label{Fig:Hall}
\end{figure}

To estimate the carrier concentration in SrMnSb$_2$, we measured Hall resistivity $\rho_{xy}$ on Batch A crystals. Figures \ref{Fig:Hall}(a) and (b) show the field dependence of $\rho_{xx}$ and $\rho_{xy}$ at $T=2$, 5, and 10 K. One can see that $\rho_{xy}$ oscillates above 6 T. Most importantly, the slopes of the curves are positive and persist in this manner to a maximum temperature of 340 K. The positive slope indicates that the dominant charge carriers responsible for transport are holes, similar to reports for BaMnSb$_2$ \cite{Huang2017} and SrMnSb$_2$ \cite{Liu2017}. A linear fit of $\rho_{xy}$ yields the Hall coefficient $R_{\rm H}$ through the relation $\rho_{xy}=R_HH$, and Figure \ref{Fig:Hall}(c) shows the temperature dependence of $R_{\rm H}$. $R_{\rm H}$ only slightly increases as the temperature is lowered to 50 K, below which it increases more abruptly. We find $R_H\approx 2.0\times10^{-7}$ m$^3$/C at $T=300$ K. For comparison, at $T=300$ K, $R_H=1.1\times10^{-10}$ m$^3$/C for optimally doped (Ba$_{1-x}$K$_x$)Fe$_2$As$_2$ ($x=0.39$) superconductors \cite{Liu2014} and $R_H=2.1\times10^{-8}$ m$^3$/C for polycrystalline Sb film with a thickness of 120 nm \cite{Volklin1990}.

Within a single band model, the carrier density can be calculated with the relation $R_H=1/pe$, where $p$ corresponds to hole concentration and $e$ is the elementary charge $1.6\times10^{-19}$ C. Therefore, $p=3.1\times10^{20}$ cm$^{-3}$ at $T=300$ K for SrMnSb$_2$. For BaMnSb$_2$ , $p=1.53\times 10^{19}$ cm$^{-3}$ at $T=100$ K \cite{Huang2017}. Taking the resistivity $\rho_{xx}=8.1\times 10^{-4}$ $\Omega$cm, one can calculate to the Hall mobility $\mu_H\approx R_H/\rho_{xx}$  as $2.5\times 10^2$ cm$^2$V$^{-1}$s$^{-1}$ at $T=300$ K. Thus in a single band model, our determination of $R_{\rm H}$ simply reflects the low charge carrier density expected for a semimetal.

\subsection{Transmission electron microscopy (TEM)}

\begin{figure}
\includegraphics[width=3.4 in]{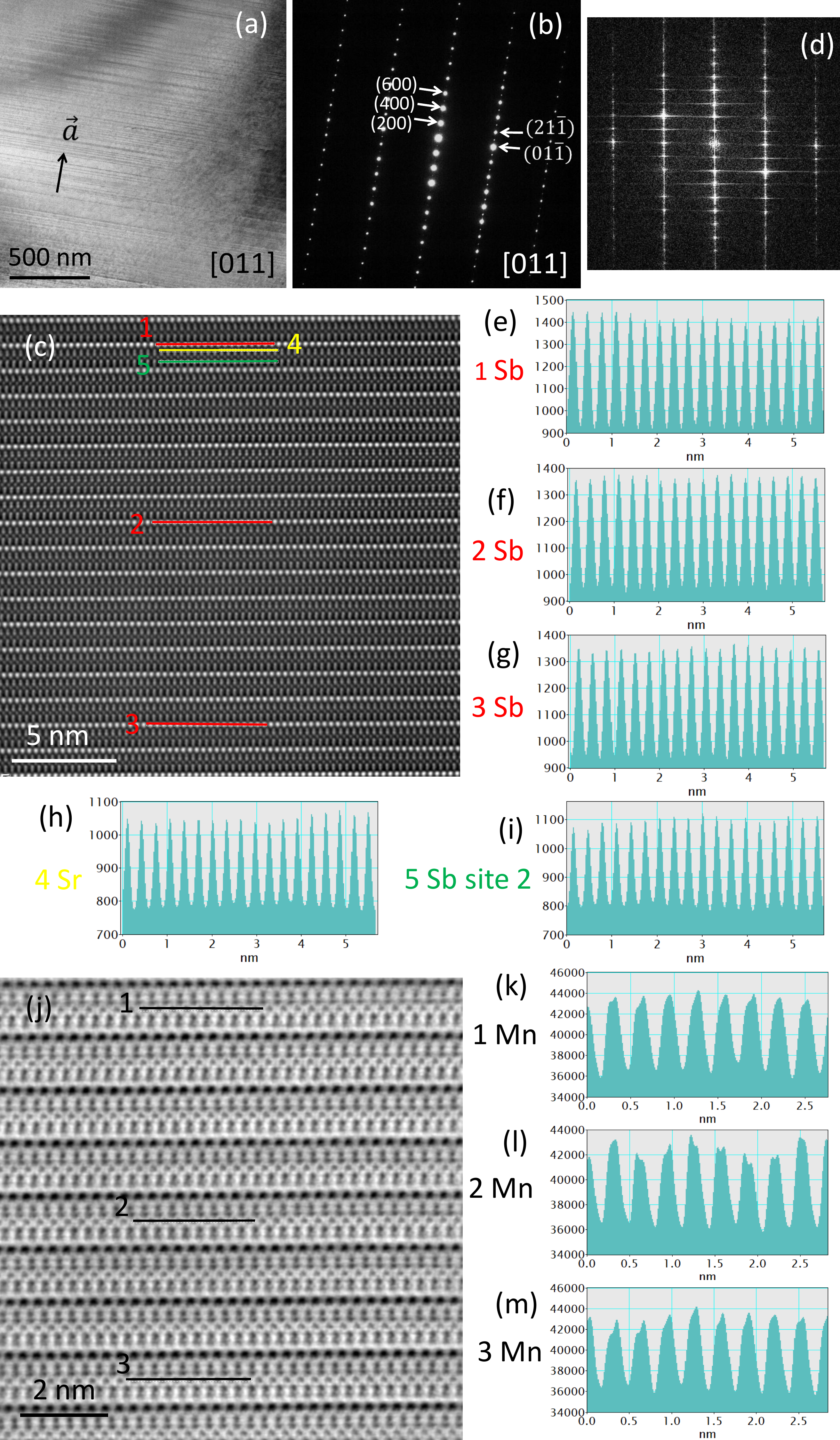}
\caption{(Color online) (a) TEM bright-field image of the [0,1,1] cross section of Batch A SrMnSb$_2$ single crystal in which stacking faults are observed. (b) SAED pattern along [0,1,1] zone axis. (c) High resolution HAADF-STEM image displaying a layer structure. (d) FFT of (c). (e)-(g) Intensity line profiles Sb columns extracted from line 1-3 in (c). (h)-(i) Intensity line profiles of Sr and Sb columns extracted from line 4 and 5 in (c), respectively. (j) High resolution ABF-STEM image of SrMnSb$_2$ single crystal. (k)-(m) Intensity line profiles of Mn columns extracted from line 1-3 in (j). }
\label{Fig:TEM1}
\end{figure}

To address the possibility of impurity phases or the presence of ordered vacancies, we performed atomic resolution TEM measurements. Figure \ref{Fig:TEM1}(a) shows a bright-field image of Batch A crystal along the [0,1,1] zone axis, with the corresponding selected area electron diffraction (SAED) pattern shown in Fig. \ref{Fig:TEM1}(b). Sharp lines perpendicular to the \textit{\textbf{a}} axis are clearly observed in the bright-field image, indicating the existence of stacking faults in the layered crystal structure. Figure \ref{Fig:TEM1}(c) shows a high resolution HAADF-STEM image, in which the layered structure is clearly identified. The FFT pattern in Fig. \ref{Fig:TEM1}(d) shows vertical streaks, again revealing stacking faults in the \textit{\textbf{bc}} plane.

\begin{figure}
\includegraphics[width=3.4 in]{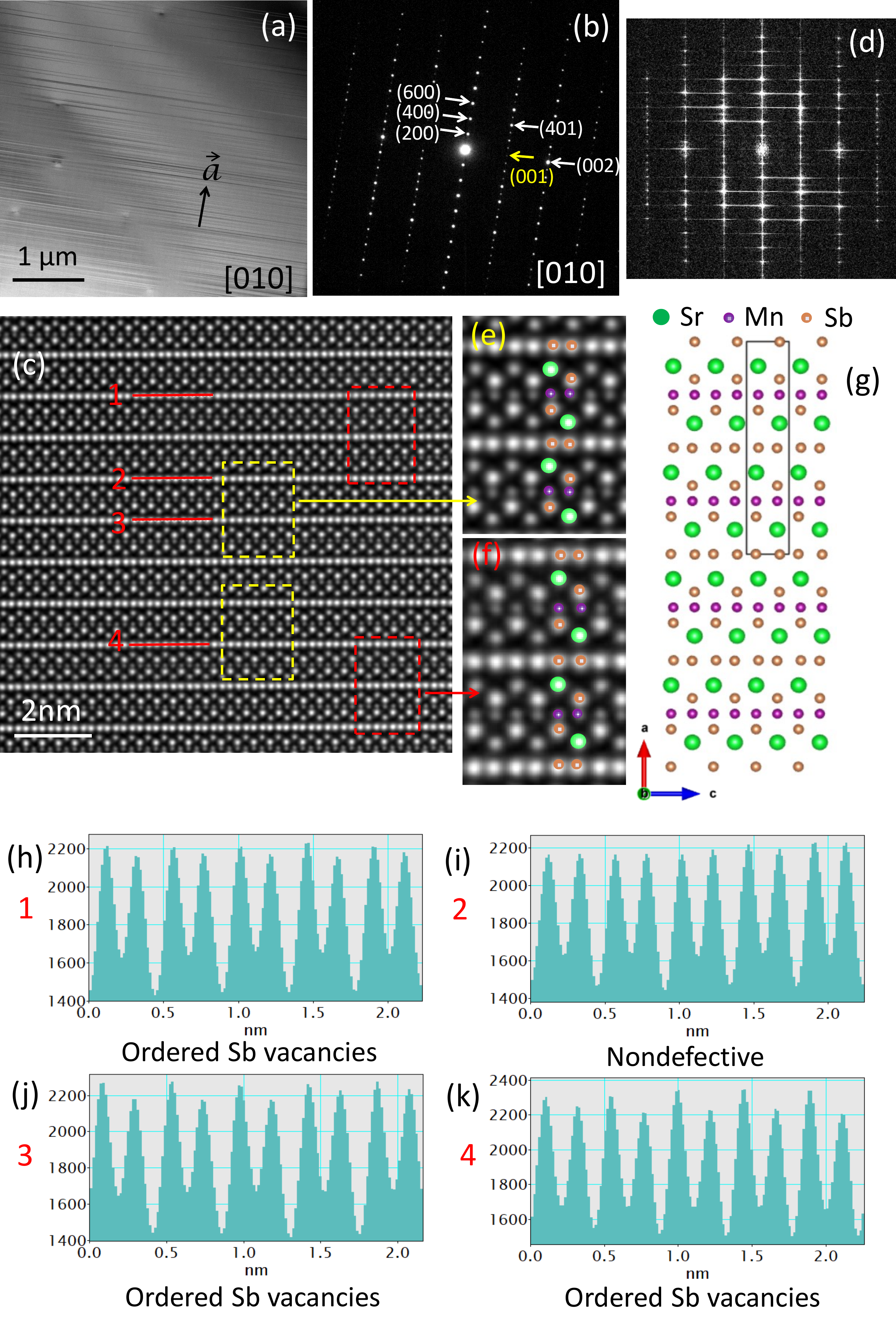}
\caption{(Color online) (a) TEM bright-field image of Batch A SrMnSb$_2$ single crystal along [0,1,0] zone axis. The sharp lines evidence the existence of stacking faults. (b) SAED pattern along [0,1,0] zone axis. The yellow index (0,0,1) is a forbidden spot in the space group $Pnma$ (62). (c) Two types of Sb atom arrangements are observed in high resolution HAADF-STEM image. Red and yellow frames indicate the normal regions and defective regions with stacking faults. (e)-(f) Magnified images from the yellow and red frames in (c). (g) Crystal structure of SrMnSb$_2$. Black line indicates a unit cell. Short and long red double arrows indicate the Sb atom pairs with narrow and wide distances. (h)-(k) Intensity line profiles of Sb columns extracted from lines 1-4 in (c). In (h), (j) and (k), one can see the strong-weak modulation in the intensity, suggesting the existence of ordered Sb vacancies; whereas no obvious variation in the intensity can be observed in (i), suggesting the layer is non-defective.}
\label{Fig:TEM2}
\end{figure}

Figures \ref{Fig:TEM1}(e)-(i) show profiles of the intensity of atomic columns, extracted from lines 1-5 as marked in Fig. \ref{Fig:TEM1}(c). Lines 1-3 (panel e-g) are associated with the Sb layers at site 1; line 4 (panel h) is for the Sr layer; line 5 (panel i) is for the Sb layer at site 2. According to Refs. \cite{Farhan2014,Huang2017}, Dirac dispersion is derived from Sb layers at site 1, i.e. Sb layers sandwiched by A (Sr) layers. In other words, Sb atoms at site 2 that sandwich the Mn atoms do not give rise to Dirac dispersion [See the crystal structure diagram in the inset of Fig. \ref{Fig:Crystals}(c) and Fig. \ref{Fig:TEM2}(g).] Our TEM analysis shows random fluctuations in the intensity of the atomic columns, which suggests no chemical variation or elemental vacancies for the [0,1,1] viewing direction. Similarly, we checked the Mn layers in an ABF-STEM image shown in Fig. \ref{Fig:TEM1}(j), which has higher sensitivity for light elements compared to HAADF (note that the atomic columns render  dark contrast in ABF). The intensity profiles for lines 1-3 are displayed in Fig. \ref{Fig:TEM1}(k)-(m). We conclude that there is no evidence of Mn deficiency in this sample since no obvious variation in the intensity can be seen.

In order to get a complete picture, the cross section for the [0,1,0] zone axis was also prepared. The representative TEM bright-field image and SAED are shown in Figs. \ref{Fig:TEM2}(a) and (b), respectively. Analogous to the previous image for [0,1,1],  sharp lines are clearly observed perpendicular to the \textit{\textbf{a}} axis, suggesting the existence of the stacking faults in \textit{\textbf{bc}} planes. From the SAED patterns along the [0,1,1] [Fig. \ref{Fig:TEM1}(b)] and [0,1,0] [Fig. \ref{Fig:TEM2}(b)] zone axes, the lattice parameters of the crystal were calculated as $a=23.30$, $b=4.403$, and $c=4.465$ {\AA}. Note that weak but clear diffraction spots at (0,0,$\pm$1), (0,0,$\pm$3) are observed in the SAED pattern along [0,1,0], which are forbidden for the $Pnma$ space group. These spots also appear in the FFT pattern shown in Fig. \ref{Fig:TEM2}(d), thus excluding the possibility of multiple scattering diffraction and suggesting the presence of certain symmetry-breaking features along \textit{\textbf{c}}. To unravel the origin of these superlattice spots, we examined the atomic layers in a high resolution HAADF-STEM image shown in Fig. \ref{Fig:TEM2}(c). Figures \ref{Fig:TEM2}(h)-(k) show the intensity profiles of Sb columns extracted from lines 1-4 in Fig. \ref{Fig:TEM2}(c). It can be clearly seen that the intensity of Sb columns shows an alternating strong-weak modulation in Figs. \ref{Fig:TEM2}(h), (j), and (k), suggesting the existence of ordered Sb vacancies in Sb layers at site 1. Therefore these ordered Sb vacancies are likely responsible for the superlattice spots, e.g., (0,0,1), (0,0,3), etc. Note that these ordered vacancies do not appear in all the Sb layers, as those along line 2 (Fig. \ref{Fig:TEM2}(i)) seem to be non-defective. It is worth mentioning that we also checked Sr and Mn layers (results not shown), and any variations in the intensity of the atomic columns were not evident.

For comparison, we discuss these results in relation to similar studies of the BaZnBi$_2$ compound \cite{Zhao2018}. BaZnBi$_2$ has a space group $I4/mmm$, and \textit{\textbf{c}} axis is perpendicular to the Zn layers. In this symmetry, (0,0,$2l+1$) peaks are not allowed; however, they appear in the XRD pattern. Zhao et al. employed atomic resolution scanning transmission electron microscopy (STEM) to directly probe the modulated structure of BaZnBi$_2$ in real space \cite{Zhao2018} and found that  the ordering of Zn vacancies appears in every second Zn layer in [1,1,0] cross section. The Zn vacancy ordering has a local character and is not uniform in BaZnBi$_2$. This local Zn vacancy ordering causes large differences in the physical properties between BaZnBi$_2$ and SrZnBi$_2$. In our case, we observe coexistence of normal and Sb-vacancy ordered regions in [0,1,0] cross section.

The high resolution image in Fig. \ref{Fig:TEM2}(c) also reveals the atomic structure of the stacking faults. Two different atomic arrangements are noticed in this image, as indicated by the yellow and red boxes. Magnified images from a yellow box and a red box are present in Figs. \ref{Fig:TEM2}(e)-(f), respectively, with Sr, Mn, and Sb atoms overlaid on each site. The atomic structure model of the pristine SrMnSb$_2$ crystal projected along [0,1,0] direction is shown in Fig. \ref{Fig:TEM2}(g), where the black box indicates the unit cell. Note that the distance between the two Sb atoms in the middle of the unit cell is narrower than that of the two Sb atoms at the bottom (as indicated by the short and long double arrows), and this sequence repeats along \textit{\textbf{a}} axis. The atomic arrangement shown in Fig. \ref{Fig:TEM2}(f) matches exactly this sequence of narrow-wide-narrow Sb pairs. However, there is a shift of the Sb layer as shown in Fig. \ref{Fig:TEM2}(e). As a result, we see narrow-narrow or wide-wide modulation of Sb pairs along \textit{\textbf{a}} axis (stacking faults).

Stacking faults have also been observed in TaAs single crystals \cite{Besara2016} for which electronic structure calculations reveal that the position of the Weyl points relative to the Fermi level shift with the introduction of vacancies and/or stacking faults. In the case of vacancies, the FS becomes considerably altered, while the effect of stacking faults on the electronic structure is to allow the Weyl pockets to remain close to the FS. The observation of quantum oscillations in the defective crystals such as TaAs, BaZnBi$_2$, and SrMnSb$_2$ suggests the robustness of the quantum phenomena in these materials to stacking faults \cite{Zhao2018,Besara2016}.

\begin{figure}
\includegraphics[width=3.4 in]{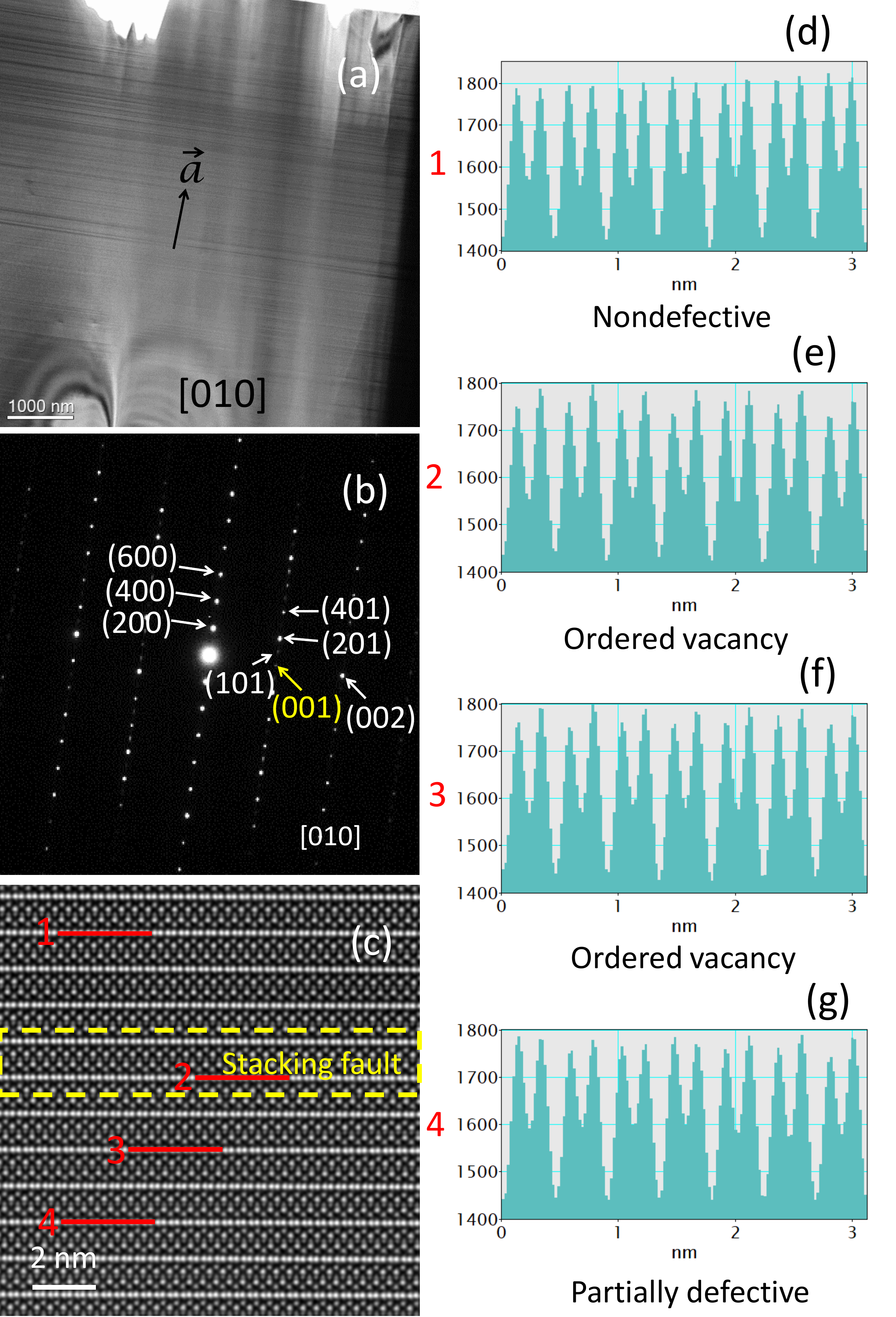}
\caption{(Color online) (a) TEM bright-field image shows the contrast of stacking faults as straight lines in Batch B crystal viewed along [0,1,0] zone axis, which are perpendicular to \textit{\textbf{a}} axis. (b) SAED pattern along [0,1,0] zone axis. The yellow index (0,0,1) is a forbidden diffraction spot in the space group $Pnma$ (62). (c) Two types of Sb atom arrangements are observed in high resolution HAADF-STEM image. Yellow frame indicates the defective regions with stacking faults. (d)-(g) Intensity profiles of Sb columns extracted from lines 1-4 in (c). In (e) and (f), one can see the strong-weak modulation in the intensity, suggesting the existence of ordered Sb vacancies.}
\label{Fig:TEM-B}
\end{figure}

Figure \ref{Fig:TEM-B} displays the TEM results on the Batch B crystal that was viewed along [0,1,0] zone axis. Sharp lines were observed in the bright-field image shown in Fig. \ref{Fig:TEM-B}(a), suggesting the existence of stacking faults. It should be emphasized that no inclusions or impurities are observed. SAED pattern shown in Fig. \ref{Fig:TEM-B}(b) presents the (0,0,1) diffraction spot, which is a forbidden spot in the space group $Pnma$ (62). A high-resolution HAADF-STEM image is shown in Fig. \ref{Fig:TEM-B}(c), in which the defective (stacking faults) region was marked by the yellow dashed box. Figures \ref{Fig:TEM-B}(d)-(g) show the intensity profile along the lines indicated in Fig. \ref{Fig:TEM-B}(c). The presence of forbidden spot (0,0,1) and strong-weak modulation in the intensity evidence the existence of the ordered Sb vacancies. The stacking faults are caused by a shift of Sb layers, as observed in the Batch A crystal (See Fig. \ref{Fig:TEM2}). Therefore, Batch A and B crystals possess the same microstructural features.

\begin{figure}
\includegraphics[width=3.4 in]{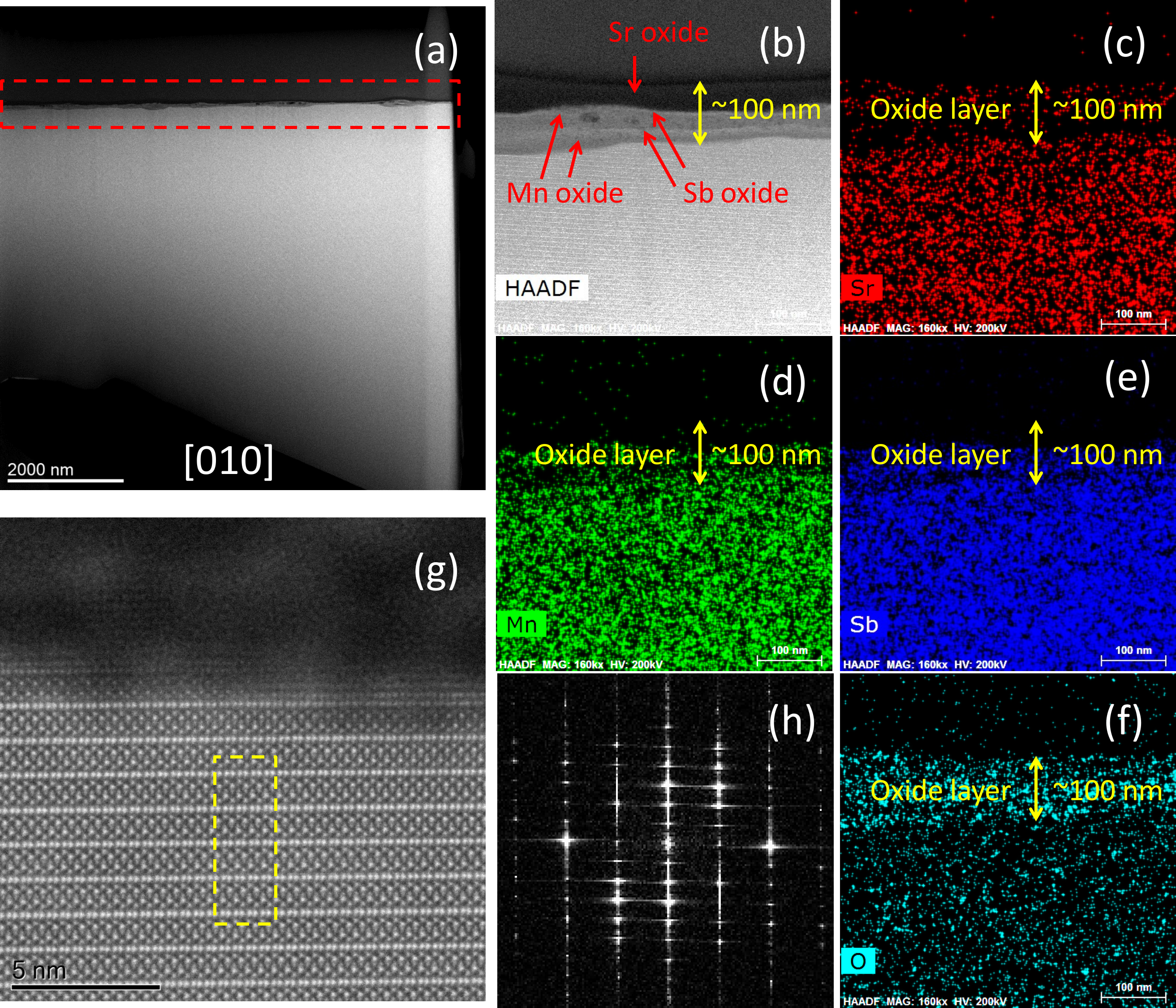}
\caption{(Color online) (a) HAADF-STEM image of cross section of SrMnSb$_2$ single crystals along [0,1,0] zone axis. The crystal was used for high-temperature susceptibility measurement. Red frame indicates the oxide layer. (b) The HAADF image shows Sr oxide, Mn oxide and Sb oxide layers with a total thickness of 100 nm. (c)-(f) Element mappings show the distributions of Sr, Mn, Sb, and O elements in the oxidized surface of single crystal. (g) Stacking faults are observed beneath the oxidized surface, as indicated by the yellow frame. (h) The streaks in SAED pattern along [0,1,0] zone axis, which evidences the stacking faults beneath the oxidized surface.}
\label{Fig:TEM3}
\end{figure}

Figure \ref{Fig:TEM3}(a) shows a low-magnification cross-sectional HAADF-STEM image of SrMnSb$_2$ single crystal along [0,1,0] zone axis. This crystal has been used in our high-temperature susceptibility measurement (see Fig. \ref{Fig:Susce3}). It is found that the surface changes from a metallic luster to a dark black color after the measurement. TEM measurements reveal that the surface of the single crystal is oxidized, as indicated by the red frame in Fig. \ref{Fig:TEM3}(a). The thickness of this oxidized layer is around 100 nm in a close view shown in Fig. \ref{Fig:TEM3}(b). Figures  \ref{Fig:TEM3}(c)-(f) show the element mappings of Sr, Mn, Sb, and O for the oxidized surface layer. Interestingly, beneath the oxidized layer, the stacking faults and ordered Sb vacancies are clearly observed, as shown in Figs. \ref{Fig:TEM3}(g)-(h). More stacking faults are observed in the region close to the oxidized layer, compared to the region away from the surface. The EDS mappings confirm that the surface of the single crystal is oxidized. It should be emphasized that there are not any inclusions observed in the cross section. Thus, the FM signal likely results from this oxidized surface. It should be pointed out that Figs. \ref{Fig:TEM3}(b)-(f) reveal that Sr oxide tends to aggregate into the outermost layer while Mn and Sb oxides form the sandwich structure. Remember that Sr$_{1-y}$Mn$_{1-z}$Sb$_2$ samples with larger Sr deficiency $y \sim 0.08$ and $z \sim 0.02$ display stronger FM behavior in Ref. \cite{Liu2017}. Our TEM results can help understand a wide spread of Sr and Mn deficiency in different crystals from the same growth batch. However, the element mappings cannot tell the existence of a MnSb phase in this sandwich structure. We assume that the surface degradation could lead to the formation of MnSb impurity phase or related compounds because MnSb is a strong ferromagnet with $T_C$ that varies with stoichiometry between 363 and 587 K \cite{Teramoto1968,Ryzhkovskii1990}. We have performed XRD measurement on single-crystal crushed powder after exposing the powder in the air. Unfortunately, as being exposed to the air, the powder immediately became red and released heat. It turned black after it was completely cooled down to room temperature. Neither a SrMnSb$_2$ nor MnSb phase survived in such a strong reaction.

\section{Conclusions}

We have grown SrMnSb$_2$ single crystals by the self flux method at different growth conditions. Susceptibility measurements show that two batches of crystals undergo AFM transition at the $T_{\rm N}=295(3)$ K. However one of the conditions yields crystals with a larger saturated magnetization that does not show anisotropy, which suggests it come from a secondary phase. The susceptibility measurement above $T_{\rm N}$ is characteristic of 2D magnetic systems where the susceptibility increases as 2D correlations develop above the 3D AFM transition. Neutron diffraction of a single crystal of SrMnSb$_2$ confirms C-type AFM order below $T_{\rm N}$ $= 295(2)$ K. XMCD measurements of a single crystal at the $L_{2,3}$ edge of Mn show a signal that is likely due to the induced canting of AFM moments under a magnetic field, namely increased XMCD signal with increased field, and puts a low limit on the canting of less than 0.01 $\mu_{\textrm{B}}$/Mn.

Magnetization and transport measurements show dHvA and SdH oscillations. The main components of these oscillations (in dHvA and SdH effects) versus inverse applied magnetic field all converge to the same FFT frequency that within error are the same for single crystals from both growth conditions. Hall effect measurements confirm that SrMnSb$_2$ is a hole dominated semimetal. Our TEM study reveals the existence of stacking faults, in both growth batches, caused by a shift of Sb atom layers and ordered Sb vacancies.

More importantly, the susceptibility at high temperatures exhibits a sharp increase that vanishes at about 600 K. Subsequent cooling from 600 K to 300 K of the same crystal shows a clear FM transition at $\sim 580 $ K that indicates a chemical transformation occurs above 500 K. This transformation likely produces a minute amount of FM impurity phase in the crystal. Neutron powder diffraction on crushed crystals at high temperatures confirm the chemical transformation observed in the susceptibility, showing formation of SrO by heating the sample under moderate vacuum above 500 K, which may create MnSb impurities. TEM measurements reveal that SrMnSb$_2$ single crystals are slightly oxidized when preparing any measurements in  air, which may display a FM transition at high-temperature in susceptibility measurements. The surface degradation could be accelerated by slightly heating, causing an enhanced FM component, which may be misunderstood as an intrinsic phenomenon. 

\begin{acknowledgments}
Work at the Ames Laboratory was supported by the Department of Energy, Basic Energy Sciences, Division of Materials Sciences \& Engineering, under Contract No. DE-AC02-07CH11358.  This research used resources of the Advanced Photon Source, a U.S. Department of Energy (DOE) Office of Science User Facility operated for the DOE Office of Science by Argonne National Laboratory under Contract No. DE-AC02-06CH11357. The work also used resources at the High Flux Isotope Reactor, a DOE Office of Science User Facility operated by the Oak Ridge National Laboratory.
\end{acknowledgments}


\begin{thebibliography}{99}
\bibitem{Geim2007} A. K. Geim and K. S. Novoselov, The rise of graphene, Nature Materials {\bf 6}, 183 (2007).

\bibitem{Wehling2014} T. O. Wehling, A. M. Black-Schaffer, A.V. Balatsky, Dirac materials, Advances in Physics {\bf 63}, 1 (2014).

\bibitem{Nielsen1981} H. B. Nielsen and N. Ninomiya, Absence of neutrinos on a lattice: (I). Proof by homotopy theory, Nucl. Phys. B {\bf 185}, 20-40 (1981); Absence of neutrinos on a lattice: (II). Intuitive topological proof, Nucl. Phys. B {\bf 193}, 173-194 (1981).

\bibitem{Wan2011} X. Wan, A. M. Turner, A. Vishwanath, and S. Y. Savrasov, Topological semimetal and Fermi-arc surface states in the electronic structure of pyrochlore iridates, Phys. Rev. B {\bf 83}, 205101 (2011).

\bibitem{Huang2015} S.-M. Huang, S.-Y. Xu, I. Belopolski, C.-C. Lee, G. Chang, B. Wang, N. Alidoust, G. Bian, M. Neupane, C. Zhang, S. Jia, A. Bansil, H. Lin, and M. Z. Hasan, A Weyl Fermion semimetal with surface Fermi arcs in the transition metal monopnictide TaAs class, Nat. Commun. {\bf 6}, 7373 (2015).

\bibitem{Xu2015} S.-Y. Xu, I. Belopolski, N. Alidoust, M. Neupane, G. Bian, C. Zhang, R. Sankar, G. Chang, Z. Yuan, C.-C. Lee, S.-M. Huang, H. Zheng, J. Ma, D. S. Sanchez, B. Wang, A. Bansil, F. Chou, P. P. Shibayev, H. Lin, S. Jia, and M. Z. Hasan, Discovery of a Weyl fermion semimetal and topological Fermi arcs, Science {\bf 349}, 613 (2015).

\bibitem{Lv2015} B. Q. Lv, N. Xu, H. M. Weng, J. Z. Ma, P. Richard, X. C. Huang, L. X. Zhao, G. F. Chen, C. E. Matt, F. Bisti, V. N. Strocov, J. Mesot, Z. Fang, X. Dai, T. Qian, M. Shi, and H. Ding, Observation of Weyl nodes in TaAs, Nat. Phys. {\bf 11}, 724 (2015).

\bibitem{Yang2015} L. X. Yang, Z. K. Liu, Y. Sun, H. Peng, H. F. Yang, T. Zhang, B. Zhou, Y. Zhang, Y. F. Guo, M. Rahn, D. Prabhakaran, Z. Hussain, S.-K. Mo, C. Felser, B. Yan and Y. L. Chen, Weyl semimetal phase in the non-centrosymmetric compound TaAs, Nat. Phys. {\bf 11}, 728 (2015).

\bibitem{Xu2015b} S.-Y. Xu, N. Alidoust, I. Belopolski, Z. Yuan, G. Bian, T.-R. Chang, H. Zheng, V. N. Strocov, D. S. Sanchez, G. Chang, C. Zhang, D. Mou, Y. Wu, L. Huang, C.-C. Lee, S.-M. Huang, B. Wang, A. Bansil, H.-T. Jeng, T. Neupert, A. Kaminski, H. Lin, S. Jia, and M. Z. Hasan, Discovery of a Weyl fermion state with Fermi arcs in niobium arsenide, Nat. Phys. {\bf 11}, 748-754 (2015).

\bibitem{Park2011} J. Park, G. Lee, F. Wolff-Fabris, Y. Y. Koh, M. J. Eom, Y. K. Kim, M. A. Farhan, Y. J. Jo, C. Kim, J. H. Shim, and J. S. Kim, Anisotropic Dirac Fermions in a Bi Square Net of SrMnBi$_2$, Phys. Rev. Lett. {\bf 107}, 126402 (2011).

\bibitem{Wang2012} K. Wang, D. Graf, L. Wang, H. Lei, S. W. Tozer, and C. Petrovic, Two-dimensional Dirac fermions and quantum magnetoresistance in CaMnBi$_2$, Phys. Rev. B {\bf 85}, 041101(R) (2012).

\bibitem{May2014} A. F. May, M. A. McGuire, and B. C. Sales, Effect of Eu magnetism on the electronic properties of the candidate Dirac material EuMnBi$_2$, Phys. Rev. B {\bf 90}, 075109 (2014).

\bibitem{Masuda2016} H. Masuda, H. Sakai, M. Tokunaga, Y. Yamasaki, A. Miyake, J. Shiogai, S. Nakamura, S. Awaji, A. Tsukazaki, H. Nakao, Y. Murakami, T. Arima, Y. Tokura, S. Ishiwata, Quantum Hall effect in a bulk antiferromagnet EuMnBi$_2$ with magnetically confined two-dimensional Dirac fermions, Science Advances {\bf 2}, e1501117 (2016).

\bibitem{Liu2017b} J. Y. Liu, J. Hu, D. Graf, T. Zou, M. Zhu, Y. Shi, S. Che, S. M. A. Radmanesh, C. N. Lau, L. Spinu, H. B. Cao, X. Ke, Z. Q. Mao, Unusual interlayer quantum transport behavior caused by the zeroth Landau level in YbMnBi$_2$, Nature Communications {\bf 8}, 646 (2017).

\bibitem{Yi2017} C. Yi, S. Yang, M. Yang, L. Wang, Y. Matsushita, S. Miao, Y. Jiao, J. Cheng, Y. Li, K. Yamaura, Y. Shi, and J. Luo, Large negative magnetoresistance of a nearly Dirac material: Layered antimonide EuMnSb$_2$, Phys. Rev. B {\bf 96}, 205103 (2017).

\bibitem{Feng2014} Y. Feng, Z. Wang, C. Chen, Y. Shi, Z. Xie, H. Yi, A. Liang, S. He, J. He, Y. Peng, X. Liu, Y. Liu, L. Zhao, G. Liu, X. Dong, J. Zhang, C. Chen, Z. Xu, X. Dai, Z. Fang, and X. J. Zhou, Strong Anisotropy of Dirac Cones in SrMnBi$_2$ and CaMnBi$_2$ Revealed by Angle-Resolved Photoemission Spectroscopy, Sci. Rep. {\bf 4}, 5385 (2014).

\bibitem{Farhan2014} M. A. Farhan, G. Lee, J. H. Shim, AEMnSb$_2$ (AE=Sr, Ba): a new class of Dirac materials, J. Phys.: Condens. Matter {\bf 26}, 042201 (2014).

\bibitem{Huang2017} S. Huang, J. Kim, W. A. Shelton, E. W. Plummer, and R. Jin, Nontrivial Berry phase in magnetic BaMnSb$_2$ semimetal, Proc. Natl. Acad. Sci. USA {\bf 114}, 6256 (2017).

\bibitem{Liu2016} J. Liu, J. Hu, H. Cao, Y. Zhu, A. Chuang, D. Graf, D. J. Adams, S. M. A. Radmanesh, L. Spinu, I. Chiorescu, and Z. Mao, Nearly massless Dirac fermions hosted by Sb square net in BaMnSb$_2$, Scientific Reports {\bf 6}, 30525 (2016).

\bibitem{Liu2017} J. Y. Liu, J. Hu, Q. Zhang, D. Graf, H. B. Cao, S. M. A. Radmanesh, D. J. Adams, Y. L. Zhu, G. F. Cheng, X. Liu, W. A. Phelan, J. Wei, M. Jaime, F. Balakirev, D. A. Tennant, J. F. DiTusa, I. Chiorescu, L. Spinu, and Z. Q. Mao, A magnetic topological semimetal Sr$_{1-y}$Mn$_{1-z}$Sb$_2$ ($y, z<0.1$), Nature Materials {\bf 16}, 905 (2017).

\bibitem{Ramankutty2018} S. V. Ramankutty, J. Henke, A. Schiphorst, R. Nutakki, S. Bron, G. Araizi-Kanoutas, S. K. Mishra, L. Li, Y. Huang, T. K. Kim, M. Hoesch, C. Schlueter, T.-L. Lee, A. Visser, Z. Zhong, J. van Wezel, E. van Heumen, and M. S. Golden, Electronic structure of the candidate 2D Dirac semimetal SrMnSb$_2$: a combined experimental and theoretical study, SciPost Phys. {\bf 4}, 010 (2018).

\bibitem{Okamoto1990} H. Okamoto, Manganese-Antimony Binary Phase Diagram (1990), ASM Alloy Phase Diagram Database, {\url{ https://matdata.asminternational.org/apd/index.aspx}}.

\bibitem{Brock1996a} S. L. Brock and S. M. Kauzlarich, Structure-property relationships in a series of mixed layer pnictide oxide compounds: A$_2$Mn$_3$Pn$_2$O$_2$ (A= Sr, Ba; Pn= P, As, Sb), J. Alloys Compd. {\bf 241}, 82 (1996).

\bibitem{Brock1996b} S. L. Brock, N. P. Raju, J. E. Greedan, and S. M. Kauzlarich, The magnetic structures of the mixed layer pnictide oxide compounds Sr$_2$Mn$_3$Pn$_2$O$_2$ (Pn= As, Sb), J. Alloys Compd. {\bf 237}, 9 (1996).

\bibitem{FullProf} J. Rodriguez-Carvajal, Recent advances in magnetic structure determination by neutron powder diffraction, Physica B {\bf 192}, 55 (1993).

\bibitem{Brechtel1981} E. Brechtel, G. Cordier, and H. Sch{\"{a}}fer, Neue tern{\"{a}}re erdalkali-{\"{u}}bergangselement-pnictide, Journal of the Less-Common Metals {\bf 79}, 131 (1981).

\bibitem{Lang2003} J. C. Lang in \textit{Characterization of Materials}, edited by E. N. Kaufmann (John Wiley \& Sons, New York, 2003), Second Edition.

\bibitem{Kartsovnik2004} M. V. Kartsovnik, High Magnetic Fields: A Tool for Studying Electronic Properties of Layered Organic Metals, Chem. Rev. {\bf 104}, 5737 (2004) 

\bibitem{Curely1998} J. Cur{\'e}ly and J. Rouch, Thermodynamics of the 2D-Heisenberg classical square lattice Part III. Study of the static susceptibility behaviours, Physica B {\bf 254}, 298 (1998). 

\bibitem{Vaknin1989} D. Vaknin, E. Caignol, P. K. Davies, J. E. Fischer, D. C. Johnston, and D. P. Goshorn. Antiferromagnetism in (Ca$_0.85$Sr$_0.15$)CuO$_2$, the parent of the cuprate family of superconducting compounds, Phys. Rev. B {\bf 39}, 9122 (1989).

\bibitem{Sangeetha2016} N. S. Sangeetha, A. Pandey, Z. A. Benson, and D. C. Johnston. Strong magnetic correlations to 900 K in single crystals of the trigonal antiferromagnetic insulators SrMn$_2$As$_2$ and CaMn$_2$As$_2$, Phys. Rev. B {\bf 94}, 094417 (2016).

\bibitem{An2009} J. An, A. S. Sefat, D. J. Singh, and M.-H. Du. Electronic structure and magnetism in BaMn$_2$As$_2$ and BaMn$_2$Sb$_2$, Phys. Rev. B {\bf 79}, 075120 (2009).

\bibitem{Singh2009} Y. Singh, A. Ellern, and D. C. Johnston, Magnetic, transport, and thermal properties of single crystals of the layered arsenide BaMn$_2$As$_2$, Phys. Rev. B {\bf 79}, 094519 (2009).

\bibitem{Singh2009b} Y. Singh, M. A. Green, Q. Huang, A. Kreyssig, R. J. McQueeney, D. C. Johnston, and A. I. Goldman, Magnetic order in BaMn$_2$As$_2$ from neutron diffraction measurements, Phys. Rev. B {\bf 80}, 100403(R) (2009).

\bibitem{Zhang2015} Q. Zhang, W. Tian, S. G. Peterson, K. W. Dennis, and D Vaknin, Spin reorientation and Ce-Mn coupling in antiferromagnetic oxypnictide CeMnAsO, Phys. Rev. B {\bf 91}, 064418 (2015).

\bibitem{McGuire2016} M. A. McGuire and V. O. Garlea, Short- and long-range magnetic order in LaMnAsO, Phys. Rev. B {\bf 93}, 054404 (2016).

\bibitem{Zhang2016} Q. Zhang, C. M. N. Kumar, W. Tian, K. W. Dennis, A. I. Goldman, and D. Vaknin, Structure and magnetic properties of {\it Ln}MnSbO ({\it Ln}= La and Ce), Phys. Rev. B {\bf 93}, 094413 (2016).

\bibitem{Liu2018} Y. Liu,  W. E. Straszheim, P. Das, F. Islam, T. W. Heitmann, R. J. McQueeney, and D. Vaknin, Synthesis and characterization of Ca-doped LaMnAsO, Phys. Rev. Mater. {\bf 2}, 054410 (2018).

\bibitem{Takei1963} W. J. Takei, D. E. Cox, and G. Shirane, Magnetic Structures in the MnSb-CrSb System, Phys. Rev. {\bf 129}, 2008-2018 (1963).




\bibitem{Hosur2013} P. Hosur and X. Qi, Recent developments in transport phenomena in Weyl semimetals, C. R. Physique {\bf 14}, 857 (2013).

\bibitem{Liu2014} Y. Liu and T. A. Lograsso, Crossover in the magnetic response of single-crystalline Ba$_{1-x}$K$_x$Fe$_2$As$_2$ and Lifshitz critical point evidenced by Hall effect measurements, Phys. Rev. B {\bf 90}, 224508 (2014).

\bibitem{Volklin1990} F. V{\"{o}}lklein, Galvanomagnetic and thermoelectric properties of antimony films, Thin Solid Films {\bf 191}, 1 (1990).

\bibitem{Zhao2018} K. Zhao, E. Golias, Q. H. Zhang, M. Krivenkov, A. Jesche, L. Gu, O. Rader, I. I. Mazin, and P. Gegenwart, Quantum oscillations and Dirac dispersion in the BaZnBi$_2$ semimetal guaranteed by local Zn vacancy order, Phys. Rev. B {\bf 97}, 115166 (2018).

\bibitem{Besara2016} T. Besara, D. A. Rhodes, K.-W. Chen, S. Das, Q. R. Zhang, J. Sun, B. Zeng, Y. Xin, L. Balicas, R. E. Baumbach, E. Manousakis, D. J. Singh, and T. Siegrist, Coexistence of Weyl physics and planar defects in the semimetals TaP and TaAs, Phys. Rev. B {\bf 93}, 245152 (2016).

\bibitem{Teramoto1968} I. Teramoto and A. M. J. G. Van Run, The existence region and the magnetic and electrical properties of MnSb, J. Phys. Chem. Solids {\bf 29}, 347 (1968).

\bibitem{Ryzhkovskii1990} V. M. Ryzhkovskii, N. D. Zhigadlo, I. L. Pashkovskii, Crystal structure and magnetic properties of some Mn$_2$Sb-based alloys, Cryst. Res. Technol. {\bf 25}, 165 (1990).
\end{thebibliography}
\end{document}